\documentclass[journal]{IEEEtai}

\usepackage[colorlinks,urlcolor=blue,linkcolor=blue,citecolor=blue]{hyperref}

\usepackage{color,array}

\usepackage{graphicx}

\usepackage{amsmath}
\usepackage{amsfonts}
\usepackage{graphicx}
\usepackage{multirow}

\begin{document}

\title{Unsupervised Representation Learning for 3-Dimensional Magnetic Resonance Imaging Super-Resolution with Degradation Adaptation} 

\author{Jianan Liu, Hao Li, Tao Huang, \IEEEmembership{Senior Member, IEEE}, Euijoon Ahn, Kang Han, Adeel Razi,\\ Wei Xiang, \IEEEmembership{Senior Member, IEEE}, Jinman Kim, \IEEEmembership{Member, IEEE}, and David Dagan Feng, \IEEEmembership{Life Fellow, IEEE}
\thanks{Jianan Liu and Hao Li contribute equally to the work and are co-first authors. Corresponding Authors: Hao Li and Tao Huang. }
\thanks{Jianan Liu is with Vitalent Consulting, Gothenburg, Sweden (email: jianan.liu@vitalent.se).}
\thanks{Hao Li is with the Department of Neuroradiology, University Hospital Heidelberg, Heidelberg, Germany (email: hao.li@med.uni-heidelberg.de).}
\thanks{Tao Huang and Euijoon Ahn are with the College of Science and Engineering, James Cook University, Cairns, Australia (email: tao.huang1@jcu.edu.au; euijoon.ahn@jcu.edu.au).}
\thanks{Kang Han and Wei Xiang are with the School of Computing, Engineering and Mathematical Sciences, La Trobe University, Melbourne, Australia (email: w.xiang@latrobe.edu.au; k.han@latrobe.edu.au).}
\thanks{Adeel Razi is with Turner Institute for Brain and Mental Health, School of Psychological Sciences, Monash University, Melbourne, Australia (email: adeel.razi@monash.edu).}
\thanks{Jinman Kim and David Dagan Feng are with the School of Computer Science, University of Sydney, Sydney, Australia (email: jinman.kim@sydney.edu.au; dagan.feng@sydney.edu.au).}
}

\markboth{IEEE Transactions on Artificial Intelligence, Vol. 00, No. 0, Month 2024}
{Liu \MakeLowercase{\textit{et al.}}: Unsupervised 3D MRI SR}

\maketitle

\begin{abstract}
High-resolution (HR) magnetic resonance imaging is essential in aiding doctors in their diagnoses and image-guided treatments. However, acquiring HR images can be time-consuming and costly. Consequently, deep learning-based super-resolution reconstruction (SRR) has emerged as a promising solution for generating super-resolution (SR) images from low-resolution (LR) images. Unfortunately, training such neural networks requires aligned authentic HR and LR image pairs, which are challenging to obtain due to patient movements during and between image acquisitions. While rigid movements of hard tissues can be corrected with image registration, aligning deformed soft tissues is complex, making it impractical to train neural networks with authentic HR and LR image pairs. Previous studies have focused on SRR using authentic HR images and downsampled synthetic LR images. However, the difference in degradation representations between synthetic and authentic LR images suppresses the quality of SR images reconstructed from authentic LR images. To address this issue, we propose a novel Unsupervised Degradation Adaptation Network (UDEAN). Our network consists of a degradation learning network and an SRR network. The degradation learning network downsamples HR images using the degradation representation learned from misaligned or unpaired LR images. The SRR network then learns to map the downsampled HR images to the original ones. Experimental results show that our method outperforms state-of-the-art networks with an improvement of up to 0.051/3.52 dB in SSIM/PSNR on two public datasets, thus is a promising solution to the challenges in clinical settings.
\end{abstract}

\begin{IEEEImpStatement}
Acquiring precisely aligned authentic high-resolution (HR) and low-resolution (LR) image pairs is considerably challenging, making supervised super-resolution (SR) reconstruction unfeasible in clinical settings. Therefore, unsupervised learning has emerged as a promising solution to this issue. This paper introduces an unsupervised network designed to be trained using unpaired or misaligned HR and LR images and enable the reconstruction of high-quality SR images. Additionally, a comparative analysis is conducted between our network and state-of-the-art unsupervised networks. This investigation delves into the divergent domain transfer strategies employed by these networks, exploring the mechanisms underpinning unsupervised domain transfer. Such insights have facilitated the identification of the most suitable approach for medical image processing.
\end{IEEEImpStatement}

\begin{IEEEkeywords}
3D Super-Resolution, Degradation Adaptation, Geometric Deformation, Magnetic Resonance Imaging, Unsupervised Learning.
\end{IEEEkeywords}

\section{Introduction}
\label{sec:introduction}

\IEEEPARstart{H}{igh} resolution (HR) magnetic resonance imaging (MRI) provides abundant soft tissue contrast and detailed anatomical structures, which help doctors with accurate diagnosis and image-guided treatment. However, the acquisition of HR images is highly time-consuming and costly. Prolonged acquisition time also leads to considerable patient discomfort, while motion artifacts are inevitable. Therefore, a reduction in the acquisition time of MRI images is highly demanded, and deep learning-based technology has been proposed to tackle this issue in recent years \cite{Wang_2016_ISBI,Zhao2019_TIP_MRISR_CSN,Zheng2021_CVPR_MRISR_SERAN,Pham2017_ISBI_MRISR_DenseNet,Pham2019_ReCNN,Chen2018_ISBI_MRISR_DenseNet,Chen2018_MICCAI_MRISR_GAN,Sui2020_MICCAI_MRISR_Gradient_Guidance,Lyu_TMI_MCSR_2020,Feng2021_MICCAI_MRISR_Multi_Contrast}.

Advancements in deep learning have facilitated the widespread adoption of super-resolution (SR) techniques across various computer vision applications, each with unique prerequisites and objectives. For instance, SR methods applied to real-world visual images and face hallucinations prioritize visual effect and perceptual performance \cite{Supervised_SR_CV_Image_1,Supervised_SR_CV_Image_2,Supervised_SR_CV_Image_3,SR_Hallucination_Image_1,SR_Hallucination_Image_2}. Conversely, applications such as remote sensing images \cite{Diffusion_for_SR_Remote_Sensing_Image,Blind_SR_Remote_Sensing_Image}, satellite videos \cite{SR_Remote_Sensing_Satellite_Video_1,SR_Remote_Sensing_Satellite_Video_2} and medical images \cite{Pham2017_ISBI_MRISR_DenseNet,Pham2019_ReCNN,Chen2018_MICCAI_MRISR_GAN,TS-RCAN} require accurate restoration of details via SR for further analysis. Notably, medical images, particularly MRI images, present unique challenges. Unlike remote sensing images and videos, which contain additional information like RGB channels \cite{Diffusion_for_SR_Remote_Sensing_Image,Blind_SR_Remote_Sensing_Image}, time series data \cite{SR_Remote_Sensing_Satellite_Video_1,SR_Remote_Sensing_Satellite_Video_2}, and multi-spectral information \cite{SR_HHyperspectral_Image_1, SR_HHyperspectral_Image_2, SR_HHyperspectral_Image_3}, MRI SR relies solely on voxel intensity. Furthermore, MRI SR requires three-dimensional (3D) reconstruction, adding further complexities to SR. This is different from standard remote sensing applications that typically involve two-dimensional (2D) image data. Therefore, the development of a dedicated network tailored for the MRI SR task is crucial to address these challenges and ensure high accuracy and reliability in the 3D space.

Training neural networks for supervised super-resolution reconstruction (SRR) can be optimized by utilizing paired low-resolution (LR) and HR MRI images, as reported in previous studies \cite{Zhao2019_TIP_MRISR_CSN,Zheng2021_CVPR_MRISR_SERAN,Pham2017_ISBI_MRISR_DenseNet,Pham2019_ReCNN,Chen2018_ISBI_MRISR_DenseNet,Chen2018_MICCAI_MRISR_GAN,Sui2020_MICCAI_MRISR_Gradient_Guidance,Lyu_TMI_MCSR_2020,Feng2021_MICCAI_MRISR_Multi_Contrast}. However, obtaining paired authentic HR and LR images - real images acquired using MRI scanners - remains a challenge in clinical settings. Besides, when the authentic HR and LR MRI image pairs are acquired in a very limited number of cases, the misalignment is introduced due to participant movement during and between the image acquisition, even in the same scan session and with highly experienced participants. The misalignment incurs significant errors in the SR image generation, making it unusable in real-world clinical practice.  

Image registration has been suggested as a straightforward solution to correct the transformation between misaligned authentic HR and authentic LR images \cite{Sui2021_MICCAI_MRISR_Generative_Degradation}. However, the limited effectiveness of rigid image registration in handling non-rigid geometric deformation caused by soft tissue movement has been reported \cite{Komninos2021_MICCAI_weaksup_nonrigid_reg_cyclegan}. Recent deformable image registration algorithms have improved accuracy, achieving a dice similarity coefficient (DSC) of around 90\% \cite{Segmentation_accuracy}. Nevertheless, such accuracy still needs to be improved when considering the size of the field of view and the object.

An alternative solution is to train the network using a combination of authentic HR images and synthetic LR images, with the synthetic LR images generated through deterministic downsampling filters, such as the Gaussian blur filter \cite{Zhao2019_TIP_MRISR_CSN,Zheng2021_CVPR_MRISR_SERAN}, or $K$-space truncation from authentic HR images \cite{Pham2017_ISBI_MRISR_DenseNet,Pham2019_ReCNN,Chen2018_ISBI_MRISR_DenseNet,Chen2018_MICCAI_MRISR_GAN,Sui2020_MICCAI_MRISR_Gradient_Guidance,Lyu_TMI_MCSR_2020,Feng2021_MICCAI_MRISR_Multi_Contrast}. After training, the trained network is used to reconstruct SR images from authentic LR images acquired separately. Most previous works in MRI SRR have followed this synthetic-LR to authentic-HR routine and have designed several different networks. For example, CSN \cite{Zhao2019_TIP_MRISR_CSN} and SERAN \cite{Zheng2021_CVPR_MRISR_SERAN} used attention mechanisms to enhance feature fusion during the 2D reconstruction procedure. ReCNN \cite{Pham2017_ISBI_MRISR_DenseNet, Pham2019_ReCNN}, and DCSRN \cite{Chen2018_ISBI_MRISR_DenseNet} were the first 3D MRI SRR networks proposed. mDCSRN \cite{Chen2018_MICCAI_MRISR_GAN}, as an extension of DCSRN, further enhanced its capability with increased depth and a discriminator. deepGG \cite{Sui2020_MICCAI_MRISR_Gradient_Guidance}, SSGNN \cite{Sui2022_SSGNN}, and SMORE (3D) \cite{Zhao2021_SMORE} generated 3D isotropic SR image volume from anisotropic LR images. MCSR \cite{Lyu_TMI_MCSR_2020} and MINet \cite{Feng2021_MICCAI_MRISR_Multi_Contrast} introduced the use of HR reference with different contrast as prior information to achieve outstanding performance. TS-RCAN \cite{TS-RCAN} also achieved performance comparable to the MINet in 3D SRR with single-contrast data and significantly reduced demands on computational resources and inference time.

However, in actual clinical settings, the performance of supervised neural networks for super-resolution reconstruction (SRR) is often compromised due to degradation shifts \cite{DA_MRISR_ETHZ}. These shifts occur because of the different degradation presentations between authentic and synthetic LR images. Unsupervised SRR approaches that use only unpaired LR and HR images for training can be employed to address this issue. For example, ZSSR \cite{ZSSR} learns the degradation from the LR image to a lower-resolution image further downsampled from the LR image. The learned degradation is then used to fit the inverse process from the LR to the HR image. Similarly, SMORE (3D) \cite{Zhao2021_SMORE} follows the same approach in 3D MRI SRR. HR-Ref-ZSSR \cite{HR_Ref_ZSSR} applies this procedure to MRI image SR with HR reference provided from another modality. ZSSR-GAN \cite{ZSSR_GAN} augments the network with a discriminator to implement a similar idea. However, these approaches still suffer from the degradation shift caused by the inconsistency between the learned degradation from the LR image to the lower resolution image and the degradation from the HR image to the LR image, and their performances are still unsatisfactory.

To address the challenge of degradation shift, Xiao \emph{et al.} \cite{Blind_SR_Remote_Sensing_Image} and Wang \emph{et al.} \cite{Wang_2021_contrastive} have proposed self-supervised methods based on contrastive learning for SRR from LR images with unknown degradation. In these methods, the networks are trained with positive and negative samples to push the generated SR images toward positive samples. These methods mitigate the degradation shift between test LR images with various degradation representations and the training LR and HR data. However, building positive and negative samples increases the complexity of the training procedure. Unlike contrastive learning approaches, domain transfer techniques are less complex and more straightforward. The degradation shift could also be addressed by employing domain transfer techniques initially designed to transfer images between different domains. The most straightforward method is CycleGAN \cite{Cyclegan}, which defines the source and target domains and facilitates image transfer between them. PseudoSR \cite{Pseudo_SR} uses CycleGAN to transfer the target domain LR image to the source domain and then reconstructs the SR image in the source domain using an SR network. However, these CycleGAN-based methods only transfer images between domains in the image space rather than the latent feature space. Another unsupervised learning method, DASR \cite{DASR}, addresses the degradation shift by calculating a domain distance map. It uses a downsampling network to transfer the HR image from the source domain to the LR image from the target domain and calculate the domain distance map simultaneously. Then, guided by loss functions based on the domain distance map, DASR trains an SR network to transfer the target domain LR image back to the source domain HR image. Although DASR is a state-of-the-art method, it lacks end-to-end training, making it difficult to train the model optimally.  However, the approaches mentioned above are designed for general computer vision SR tasks. They may not be appropriate for MRI SRR.

To address the challenges posed by the lack of aligned authentic LR and HR MRI image pairs and the degradation shift, we propose an unsupervised method that can utilize misaligned or unpaired HR and LR images to train the network. The key contributions of our work are summarized below.

\begin{itemize}
    \item We introduce an end-to-end unsupervised degradation adaptation deep neural network (UDEAN), which employs a degradation adaptation (DA) mechanism to adaptively learn the degradation representation between the misaligned or unpaired authentic LR and HR MRI images in both the image space and the latent feature space. The combination of DA in both spaces significantly enhances the quality of the reconstructed SR images, surpassing the performance of DA in any single space.
    \item Our proposed DA mechanism outperforms existing domain transfer methods in the literature. Specifically, the DA mechanism transfers the HR image of the source domain to the LR image of the target domain with an unknown degradation representation, minimizing the errors in the reconstructed SR MRI images.
    \item The proposed method can be trained with either misaligned or unpaired LR and HR MRI images, making it suitable for real-world clinical settings where accurately aligned authentic LR and HR MRI image pairs are not available.
    \item Our experimental results demonstrate that our proposed method has distinct advantages over existing paired supervised solutions, achieving up to 0.022/1.49 dB improvement in SSIM/PSNR, and outperforming other unsupervised SR training methods by up to 0.051/3.52 dB in SSIM/PSNR across two diverse public datasets and two scale factors.
\end{itemize}

\section{Methodology}

\begin{figure}
\begin{center}
\includegraphics[width=240pt]{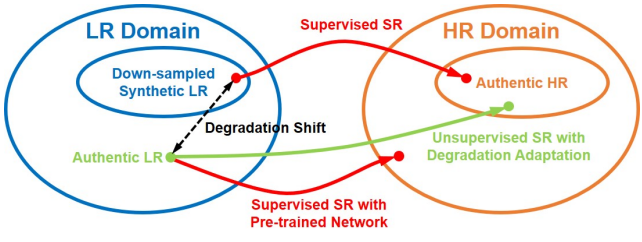}
\caption{Domain interpretation of differences between supervised and unsupervised SR. A large degradation shift exists between the SR result and desired HR image, which is caused by applying a supervised network, which is pre-trained with synthetic LR images, to authentic LR images with degradation deviating.} \label{degradation_shift}
\end{center}
\end{figure}

\begin{figure*}
\centering
\includegraphics[width=500pt]{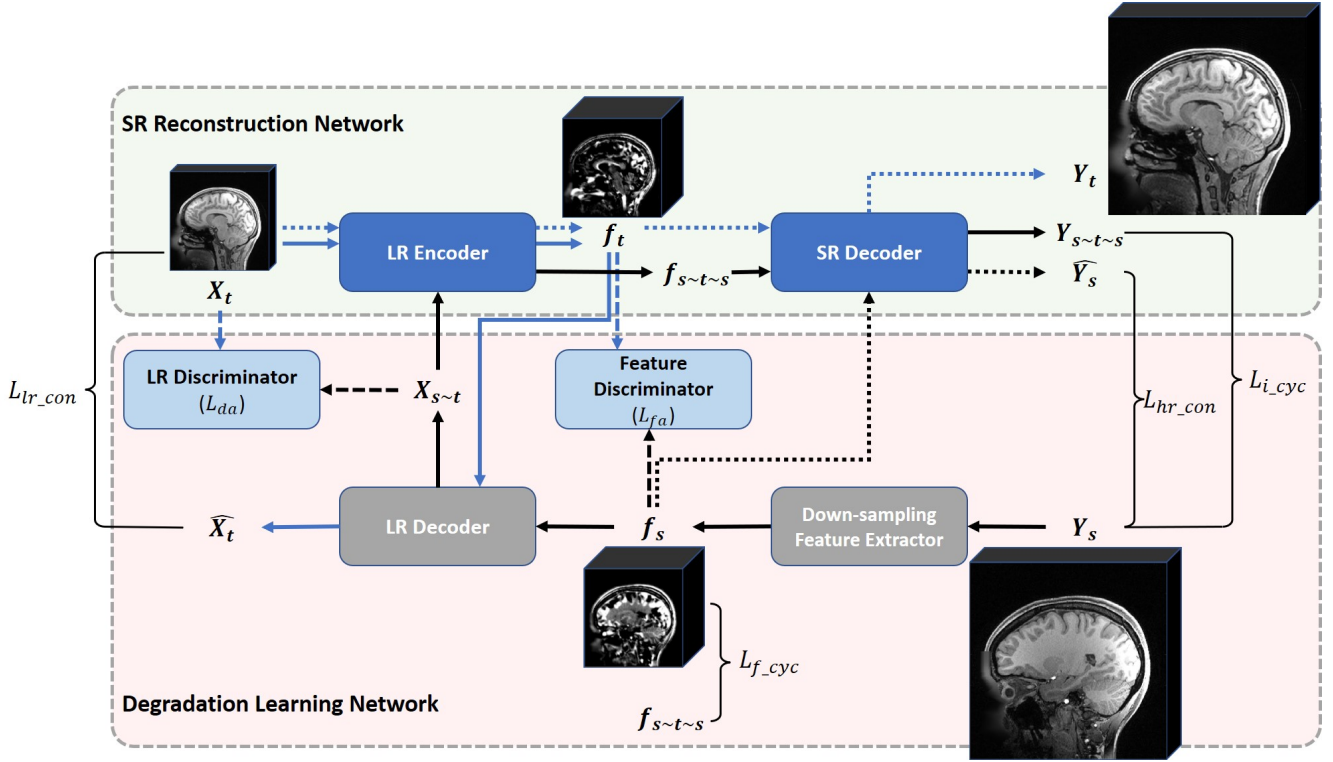}
\caption{Pipeline of the UDEAN for 3D MRI super-resolution reconstruction. The network is fed with the unpaired or misaligned source group HR image patch $\textbf{Y}_{\textbf{s}}$ and the target group LR patch $\textbf{X}_{\textbf{t}}$ in training. During the inference, only the target group LR image patch $\textbf{X}_{\textbf{t}}$ is fed in the network, and the SR image patch $\textbf{Y}_{\textbf{t}}$ is reconstructed.} \label{Fig_network}
\end{figure*}

\subsection{Degradation Shift and SR Reconstruction with Unknown Degradation}

The degradation shift, also known as the domain gap in the real-world image SRR, was initially noticed in blind SRR tasks, where the downsampling parameters of the LR images were unknown \cite{Liu_2021_survey}. In previous studies of real-world image SRR, HR images were downsampled with a bicubic algorithm or specific Gaussian blurring kernels \cite{Liu_2021_survey,Gu_2019_kernel,Xie_2021_kernel,Hui_2021_kernel,Yamac_2021_kernel,Zhang_2022_kernel,Zhang_2021_kernel,Wang_2021_contrastive,Zhang_2021_contrastive}. With these downsampling algorithms, the parameters of the algorithms for the LR images were known in both training and inference datasets. They should be identical to guarantee the performance of the trained network \cite{Liu_2021_survey}. However, the parameters of the downsampling algorithms in the real world were normally unknown and very difficult to model \cite{Liu_2021_survey}. Therefore, it is impossible to guarantee identical downsampling for the LR images in both the training dataset and the inference dataset, and the difference between the degradation of the training and inference datasets resulted in a downgraded performance of the trained network in inference\cite{Liu_2021_survey}, as shown in Fig. \ref{degradation_shift}.

Similarly, in MRI SRR, $K$-space truncation is the most widely used downsampling algorithm to generate synthetic LR images \cite{Pham2017_ISBI_MRISR_DenseNet,Pham2019_ReCNN,Chen2018_ISBI_MRISR_DenseNet,Chen2018_MICCAI_MRISR_GAN,Sui2020_MICCAI_MRISR_Gradient_Guidance,Lyu_TMI_MCSR_2020,Feng2021_MICCAI_MRISR_Multi_Contrast}. With this algorithm, an HR image is transformed into its fake $K$-space using the fast Fourier transformation (FFT). The fake $K$-space is truncated based on the downsampling factor. At last, the remained part of the fake $K$-space is transformed back to the synthetic LR image using inverse FFT (iFFT). However, the degradation shift between the synthetic and authentic LR images created in this process still remained \cite{DA_MRISR_ETHZ}. The method of $K$-space truncation is a highly simplified downsampling model, which simulates the acquisition process of LR images based on the assumption that the $K$-spaces of both the HR and the LR images were fully sampled. However, to shorten the scan time in the clinical measurements, most of the protocols for both LR and HR images acquire $K$-spaces, which are under-sampled in several different manners, including partial Fourier in phase-encoding and/or slice-encoding direction, parallel imaging, elliptical $K$-space, etc. Most of these under-sampling processes are difficult to simulate accurately since many of the key parameters are unknown. In addition, some other factors also lead to unpredictable differences between the degradation representations of synthetic and authentic LR images, such as the varied signal-to-noise ratio caused by the different signal intensity distribution of the $K$-spaces or different tuning of the transmission/receiving chain.

\begin{figure*}
\centering
\includegraphics[width=450pt]{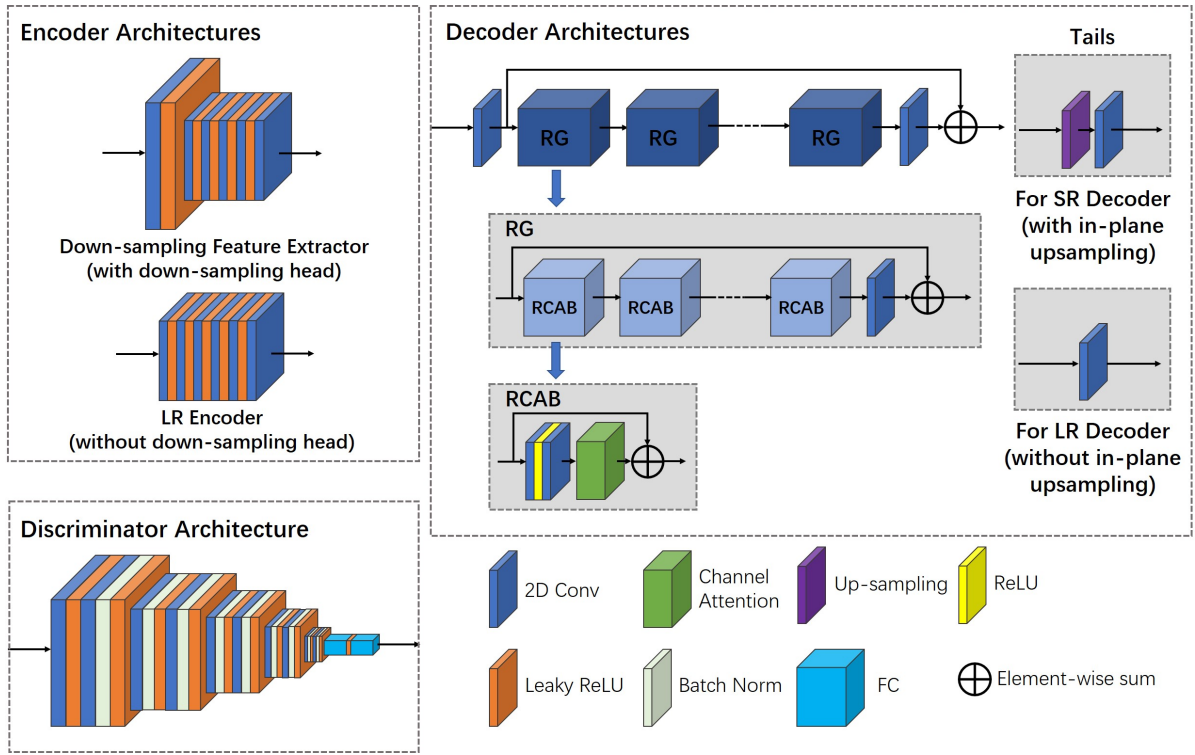}
\caption{Detailed structures of network components. The encoders are constructed with 6 convolutional layers, each followed by a Leaky ReLU layer. The decoders adopt the TS-RCAN backbone. The VGG network is employed as the discriminator.} \label{Fig_network_detail}
\end{figure*}

Therefore, unsupervised SRR can be a promising solution when the degradation shift exists. Several methods have been proposed to address this problem. These methods can be categorized into three types: the generative network-based domain transfer methods \cite{Cyclegan,Pseudo_SR,DASR}, the blurring kernel-based methods \cite{Liu_2021_survey,Gu_2019_kernel,Xie_2021_kernel,Hui_2021_kernel,Yamac_2021_kernel,Zhang_2022_kernel,Zhang_2021_kernel}, and contrastive learning \cite{Blind_SR_Remote_Sensing_Image,Wang_2021_contrastive,Zhang_2021_contrastive}. Since the LR images in the MRI were not downsampled with either bicubic or Gaussian blurring kernels, the blurring kernel-based methods are incompatible with the MRI SRR. Contrastive learning requires positive and negative training samples, which are difficult to acquire, making this method difficult to implement in the MRI SRR. Therefore, a network with degradation adaptation based on the domain transfer method is the most straightforward option and was adopted in this paper for SRR from LR images with unknown degradation.

\subsection{Network Architecture of the UDEAN}

In terms of unsupervised training, images from the source and target groups are used in the training process. The images from the target group are the LR images that will be upsampled to reconstruct the SR images. The images from the source group provide the HR images with the style that the SR images expect. As shown in Fig. \ref{Fig_network}, the HR MRI image $\textbf{Y}_{\textbf{s}}$ from the source group and the LR MRI image $\textbf{X}_{\textbf{t}}$ from the target group are misaligned or unpaired, and there is no ground truth HR image to calculate any pixel-wise or structural losses with the generated SR image. Noting that misaligned LR and HR images can be considered a special case of unpaired data, where the LR image contains similar structural information as the HR image but is not pixel-wise aligned. Therefore, the training is still not applicable in a supervised manner.

Therefore, we propose an unsupervised degradation adaptation network, UDEAN, whose architecture is shown in Fig. \ref{Fig_network}. The network contains two components: a degradation learning network and an SR reconstruction network. There are two modules in the degradation learning network. The first module is a degradation mapping module, colored gray as shown in Fig. \ref{Fig_network}. This module contains a downsampling feature extractor and an LR decoder. The second module is an identification module consisting of an LR discriminator and a feature discriminator, as shown in Fig. \ref{Fig_network} in light blue color. The SRR network includes an LR encoder and an SR decoder (colored deep blue). It is expected to reconstruct SR images with comparable image quality to the HR images of the source group from the LR images of the target group.

As indicated along the black solid line, $\textbf{Y}_{\textbf{s}}$ is fed to a learnable downsampling feature extractor to extract the downsampled feature map $f_s$ of the source group. This feature map contains a generic degradation representation from HR to LR images. Then, the downsampled LR image $X_{{s}\sim{t}}$, which contains a comparable degradation representation of the target group LR image $\textbf{X}_{\textbf{t}}$, is generated by the LR decoder. Then the feature map $f_{{s}\sim{t}\sim{s}}$ of $X_{{s}\sim{t}}$ is extracted using an LR encoder and generates the SR image $Y_{{s}\sim{t}\sim{s}}$ using an SR decoder. As the reconstructed SR image $Y_{{s}\sim{t}\sim{s}}$ and HR image in the source group $\textbf{Y}_{\textbf{s}}$ are expected to be identical, and the feature map $f_{{s}\sim{t}\sim{s}}$ and $f_s$ should likewise contain a comparable context, two cycle losses (HR image and feature cycle losses) are calculated for such purposes. Two consistency losses are also involved in making training more stable. The LR encoder and LR decoder are trained by minimizing the target group error between $\textbf{X}_{\textbf{t}}$ and the LR image $\hat{X_t}$ decoded from $f_t$ along the blue solid line. Similarly, the downsampling feature extractor and SR decoder are trained by minimizing the source group error between $\textbf{Y}_{\textbf{s}}$ and the SR image $\hat{Y_s}$ reconstructed from $f_s$ along the black dotted line.

Since the network is expected to learn the specific degradation representation from the HR images in the source group to the LR images in the target group, an LR discriminator and a feature discriminator are employed. The former is trained to distinguish between the target group LR image $\textbf{X}_{\textbf{t}}$ and the LR image $X_{{s}\sim{t}}$ reconstructed from source group feature. The second discriminator is trained to distinguish between the feature $f_t$ retrieved from the target group LR and $f_s$. The discriminators have been trained alternately with the degradation mapping module and the SRR network. The UDEAN can adapt from the source group to the specific degradation representation in the target group as long as the discriminators are deceived.

Only the trained LR encoder and SR decoder are required during the inference step, as shown along the blue dotted line in Fig. \ref{Fig_network}. The target group LR MRI image $\textbf{X}_{\textbf{t}}$ is encoded into the feature map, then the target group SR MRI image $\textbf{Y}_{\textbf{t}}$ is generated accordingly.

\subsection{Loss Functions}

In our experiments, we employed three types of loss functions, which are the L1 loss, the structural similarity (SSIM) loss \cite{Masutani2020_Radiology_SSIMLoss}\cite{SSIM} and the adversarial loss in the training of the generators:
\begin{equation}
L_1(x,y)=\frac{1}{N}\sum_{i=1}^{N}\lvert x-y\rvert
\end{equation}
\begin{equation}
L_{SSIM}(x,y)=\frac{1}{N}\sum_{i=1}^{N}\lvert 1-SSIM(x,y)^2\lvert
\end{equation}
\begin{equation}
L_{adv}(x)=\frac{1}{N}\sum_{i=1}^{N}(D(x)-1)^2
\end{equation}
where the batch size is $N$, $x$ and $y$ represent the generated SR image and ground truth HR images, and $D(\cdot)$ represents the discriminator. To stabilize the training procedure, we use the least square loss \cite{Least_square_loss} for the adversarial loss in our model instead of the negative log-likelihood \cite{Negative_log_likelihood}. 

As mentioned in the previous section, the two-cycle losses are used to guide the cross-domain restoration of $Y_{{s}\sim{t}\sim{s}}$. The image cycle loss is calculated as a weighted sum of the L1 loss and SSIM loss between $Y_{{s}\sim{t}\sim{s}}$ and $\textbf{Y}_{\textbf{s}}$, and the adversarial loss of $Y_{{s}\sim{t}\sim{s}}$. The feature cycle loss is calculated as the L1 loss between $f_{{s}\sim{t}\sim{s}}$ and $f_s$:
\begin{equation}
\begin{aligned}
L_{i\_cyc}=L_1(Y_{{s}\sim{t}\sim{s}},\textbf{Y}_{\textbf{s}})+\alpha*L_{SSIM}(Y_{{s}\sim{t}\sim{s}},\textbf{Y}_{\textbf{s}})\\+\beta*L_{adv}(Y_{{s}\sim{t}\sim{s}})
\end{aligned}
\end{equation}
\begin{equation}
L_{f\_cyc}=L_1(f_{{s}\sim{t}\sim{s}},f_s)
\end{equation}

In addition, consistency losses are used to restrict the restoration of $\hat{Y_s}$ and $\hat{X_t}$ within the source and target domain, respectively. They are calculated as a weighted sum of L1, SSIM, and adversarial loss:
\begin{equation}
\begin{aligned}
L_{hr\_con}=L_1(\hat{Y_s},\textbf{Y}_{\textbf{s}})+\alpha*L_{SSIM}(\hat{Y_s},\textbf{Y}_{\textbf{s}})\\+\beta*L_{adv}(\hat{Y_s})
\end{aligned}
\end{equation}
\begin{equation}
\begin{aligned}
L_{lr\_con}=L_1(\hat{X_t},\textbf{X}_{\textbf{t}})+\alpha*L_{SSIM}(\hat{X_t},\textbf{X}_{\textbf{t}})\\+\beta*L_{adv}(\hat{X_t})
\end{aligned}
\end{equation}
where $\alpha$ and $\beta$ are the weights of the SSIM and adversarial loss, respectively. We set $\alpha=0.5$ and $\beta=0.01$ in our experiments.

Furthermore, the DA losses in image space ($L_{da}$) and in latent feature space ($L_{fa}$), which stemmed from the LR discriminator and the feature discriminator, are used to guide the network approaching the distributions of the LR images and the features extracted from different domains:
\begin{equation}
\begin{aligned}
L_{da}(X_{{s}\sim{t}},\textbf{X}_{\textbf{t}})=\frac{1}{N}\sum_{i=1}^{N}\lvert D_{da}(X_{{s}\sim{t}})-0.5\rvert\\+\frac{1}{N}\sum_{i=1}^{N}\lvert D_{da}(\textbf{X}_{\textbf{t}})-0.5\rvert
\end{aligned}
\end{equation}
\begin{equation}
\begin{aligned}
L_{fa}(f_t, f_s)=\frac{1}{N}\sum_{i=1}^{N}\lvert D_{fa}(f_t)-0.5\rvert\\+\frac{1}{N}\sum_{i=1}^{N}\lvert D_{fa}(f_s)-0.5\rvert
\end{aligned}
\end{equation}

As a result, the end-to-end training loss for the generators in our network is defined as
\begin{equation}
\begin{aligned}
Loss_{G}=\lambda_1*L_{i\_cyc}+\lambda_2*L_{f\_cyc}+\lambda_3*L_{hr\_con}\\+\lambda_4*L_{lr\_con}+\lambda_5*L_{da}+\lambda_6*L_{fa}
\end{aligned}
\end{equation}
where $\lambda_1$ to $\lambda_6$ are the weights of the loss components. We set $\lambda_{1,3,4}=1$ and $\lambda_{2,5,6}=0.1$ in our experiments for the best performance.

For the training of the LR discriminator (LRD) and the feature discriminator (FD), we also employed the least square loss:
\begin{equation}
\begin{aligned}
L_{LRD}(\textbf{X}_{\textbf{t}}, X_{{s}\sim{t}})=\frac{1}{N}\sum_{i=1}^{N}(D_{da}(\textbf{X}_{\textbf{t}})-1)^2 \\ +\frac{1}{N}\sum_{i=1}^{N}(D_{da}(X_{{s}\sim{t}})-0)^2
\end{aligned}
\end{equation}
\begin{equation}
\begin{aligned}
L_{FD}(f_s, f_t)=\frac{1}{N}\sum_{i=1}^{N}(D_{fa}(f_s)-1)^2 \\ +\frac{1}{N}\sum_{i=1}^{N}(D_{fa}(f_t)-0)^2
\end{aligned}
\end{equation}

\subsection{Dataset and Data Pre-processing}

In this study, we used two public datasets:
\subsubsection{Human Connectome Project (HCP) dataset} The T1w images from the HCP dataset, which consists of images from 1113 healthy participants \cite{HCP}. The T1w images were acquired in the sagittal plane with 3D MPRAGE, $TR=2400ms, TE=2.14ms, TI=1000ms, FA=8, FOV=224mm, rBW=210Hz/Px$, the matrix size was $320\times320\times256$ with isotropic resolution $0.7mm$, and $\times2$ GRAPPA was activated in the phase encoding direction. To shorten the training time, we randomly selected 300 participants for our study.
\subsubsection{Brain Tumor Segmentation Challenge 2020 (BraTS) dataset} The contrast-enhanced T1w (T1CE) images from the BraTS dataset, which consists of images from 369 participants with brain tumors\cite{BraTS_1,BraTS_2,BraTS_3}. The T1CE images were acquired in the axial plane, the matrix size was $240\times240\times155$ with isotropic resolution $1.0mm$. To shorten the training time, we randomly selected 300 participants for our study.

\begin{figure*}
\centering
\includegraphics[width=\textwidth]{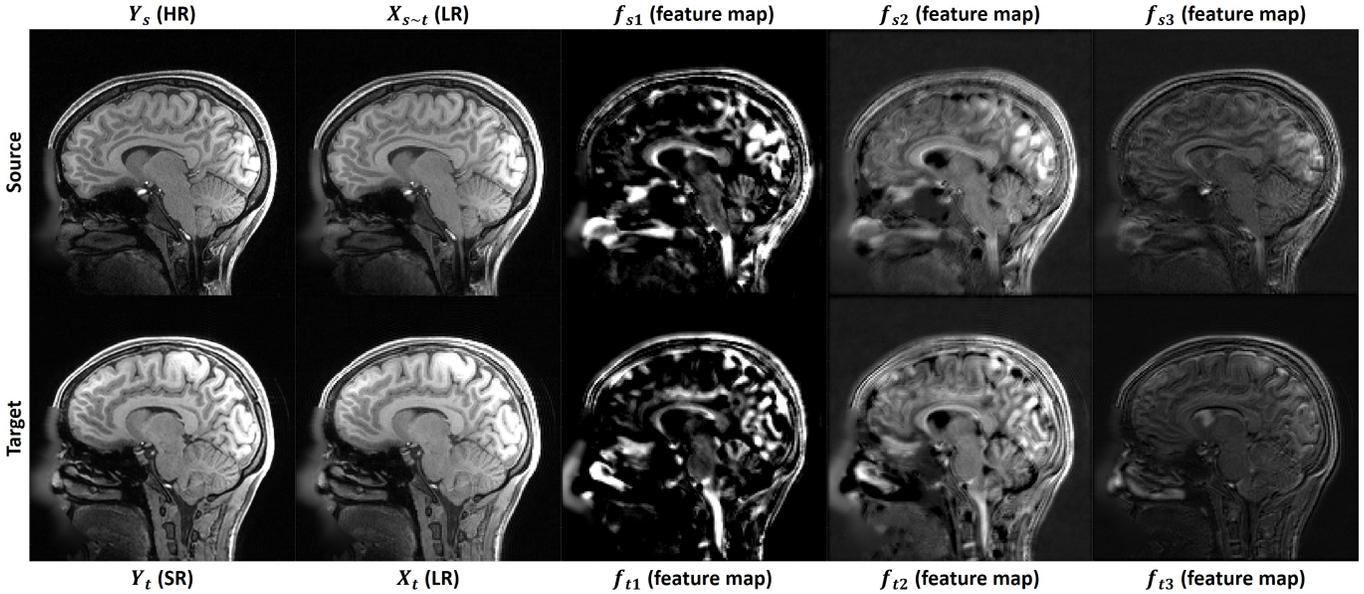}
\caption{Illustration of the degradation adaptation and feature extraction. Images with similar anatomical structures were selected from different participants in the source group (in the top row) and the target group (in the bottom row). $\textbf{Y}_{\textbf{s}}$ was the HR image from the source group, and $\textbf{Y}_{\textbf{t}}$ was the SR image reconstructed using the LR image ($X_t$) from the target group. $X_{{s}\sim{t}}$ was the LR image degraded from $\textbf{Y}_{\textbf{s}}$ by the degradation learning network, and its image quality was comparable to $X_t$. $f_{s1}$ to $f_{s3}$ and $f_{t1}$ to $f_{t3}$ were examples of the feature maps extracted from $\textbf{Y}_{\textbf{s}}$ and $X_t$, respectively. Highly consistent patterns could be found between the feature maps of the source group and the target group, revealing the effectiveness of the degradation adaptation in the latent feature space. } \label{degradation_illustration}
\end{figure*}

\begin{figure}
\centering
\includegraphics[width=250pt]{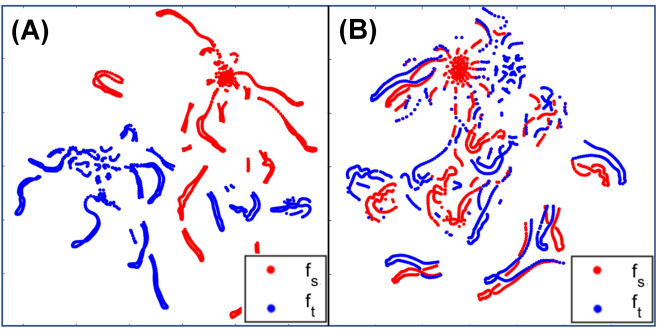}
\caption{Illustration of the degradation adaptation process using t-SNE. (A): The distributions of the feature maps from the source domain ($f_s$) and the target domain ($f_t$) were differentiable before degradation adaptation was performed. (B): After degradation adaptation, the distributions of the feature maps overlapped.} \label{DA_illustration}
\end{figure}

Our experiments involved four data groups: the source, target, validation, and test groups. The source and target groups were used to train the neural networks, the validation group was used to monitor the network performance during training, and the test group was used to evaluate the networks after training. We randomly selected 120/120/30/30 participants from the HCP dataset for the source, target, validation, and test group. The target group contains only LR images, which were downsampled from HR images by 3D $K$-space truncation with scale factors of $2\times2\times1$ and $2\times2\times2$ \cite{Zhao2019_TIP_MRISR_CSN,Zheng2021_CVPR_MRISR_SERAN,Pham2017_ISBI_MRISR_DenseNet,Pham2019_ReCNN,Chen2018_ISBI_MRISR_DenseNet,Chen2018_MICCAI_MRISR_GAN,Sui2020_MICCAI_MRISR_Gradient_Guidance,Lyu_TMI_MCSR_2020,Feng2021_MICCAI_MRISR_Multi_Contrast}. The source group contained only HR images from 120 participants and differed between the experiments with misaligned/unpaired HR and LR images. For the unpaired SRR, the participants of the source group were isolated from the other three groups; for the misaligned SRR, the source group shared the same participants with the target group, and the HR images were distorted with certain deformation patterns. 

Specifically, for the SRR from misaligned LR and HR image pairs, the participant movement between the acquisition of HR and LR images was simulated by adopting mild rigid movement and non-rigid geometric deformation to the HR images. The rigid movement was performed by random rotation of the image volumes by 0 to 2 degrees around the head-feet (H-F) axis and the left-right (L-R) axis and random translation by 0 to 2 voxels in H-F and head-feet L-R directions. The geometric deformation was achieved by randomly shrinking the whole image volumes by 0 to 2 voxels in the anterior-posterior (A-P) and L-R directions, resulting in a 0.7 to 0.9\% change in the objects' sizes. Each dataset was scaled to the range of 0 to 1, and to save computation resources, the LR and HR images were cropped into 3D patches of sizes $64\times64\times3$ and $128\times128\times3$ for the scale factor of $2\times2\times1$, and $64\times64\times3$ and $128\times128\times6$ for the scale factor of $2\times2\times2$, respectively.

\begin{table*}[ht]
\centering
\caption{Quantitative results of ablation study for MRI SRR with misaligned LR and HR images of the HCP dataset and scale factor of $2\times2\times2$.}
\label{tab1}
\setlength{\tabcolsep}{12pt}
\begin{tabular}{c|c|c|c|c|c}
\hline
     & Supervised without & Supervised with    & Unsupervised with    & Unsupervised with  & Unsupervised with  \\
     & Rigid-Registration & Rigid-Registration & DA in Feature Space & DA in Image Space  & DA in Both Spaces  \\
\hline
SSIM & $0.8939\pm0.0140$  & $0.9025\pm0.0116$  & $0.9162\pm0.0104$    & $0.9229\pm0.0093$  & $\mathbf{0.9247\pm0.0097}$  \\
PSNR & $32.7886\pm1.5546$ & $32.7823\pm1.5013$ & $33.5224\pm1.6849$   & $33.5642\pm1.7719$ & $\mathbf{34.2729\pm1.8080}$ \\
\hline
\end{tabular}
\end{table*}

\begin{figure*}
\centering
 \includegraphics[width=\textwidth]{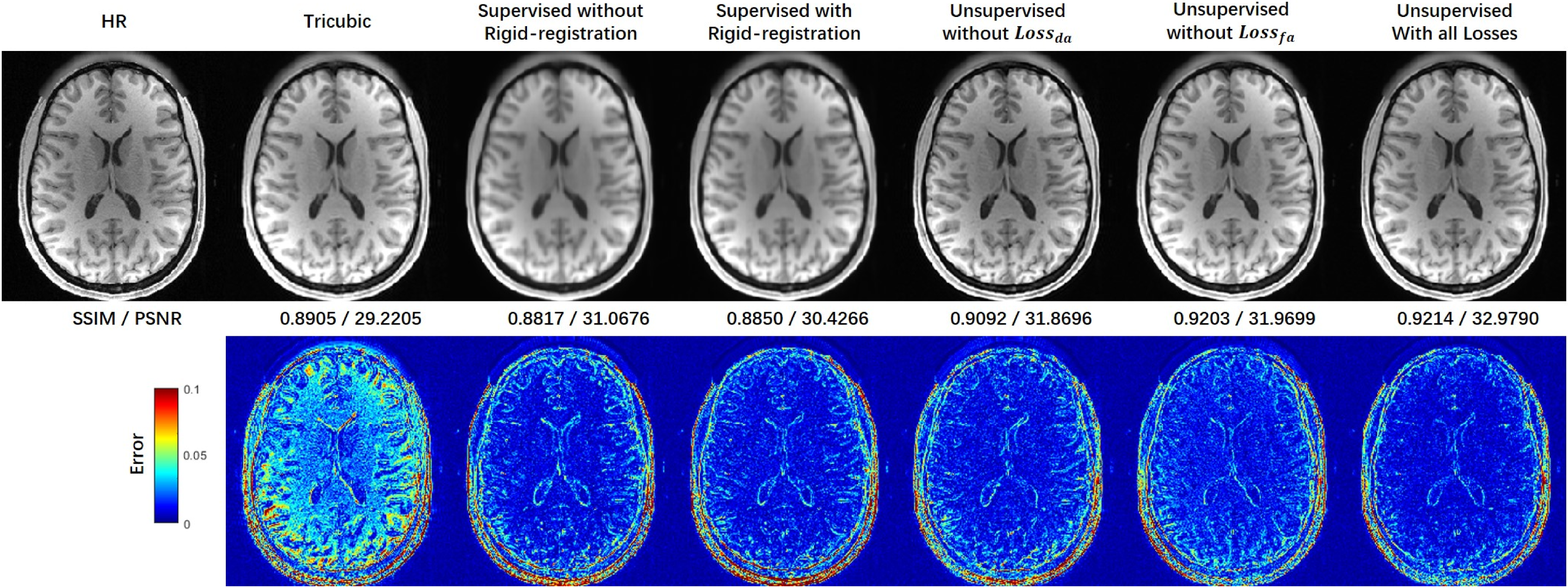}
\caption{Comparison of reconstructed SR images and error maps of supervised learning and different configurations of the UDEAN with misaligned LR and HR image pairs of the HCP dataset and the scale factor of $2\times2\times2$ in the axial plane. The tricubic interpolated image has many errors in the detailed structure. SR images reconstructed using supervised learning are highly blurry, even with a geometric deformation below 1\%. The contribution of degradation adaptation can be observed in the images with a single type of degradation adaptation and with both degradation adaptations. The absence of any degradation adaptation leads to lower accuracy of the reconstructed images.} \label{comparison_UDEAN_and_supervised_approaches}
\end{figure*}

\subsection{Implementation Details}
The detailed structures of our model components are shown in Fig. \ref{Fig_network_detail}. For the encoders, we used 6 convolution layers with Leaky ReLU between every two layers, and the downsampling feature extractor downsamples the input HR image to the same size as the LR image. We used TS-RCAN \cite{TS-RCAN} with 5 residual groups (RG) and 5 residual blocks (RCAB) in each RG as the backbone of the decoders and VGG \cite{VGGnet} as the discriminator for the UDEAN. TS-RCAN is modified from RCAN \cite{RCAN,Revisiting_RCAN} to conduct 3D MRI SR tasks with low consumption of computation resources and short inference time.

The networks were trained on a workstation equipped with an Nvidia Quadro A6000. For all deep learning experiments, we used Pytorch 1.9 as the back end. In each training batch, eight LR patches were randomly extracted as inputs. We trained our model for 30 epochs using the ADAM optimizer with $\beta_1 = 0.9$, $\beta_2 = 0.99$, and $\epsilon= 10^{-8}$, and a Cosine-decay learning rate was applied from $10^{-4}$ to $10^{-8}$. The image pre-processing of downsampling, deformation, and cropping, the post-processing, and metrics calculation were performed with Matlab 2020a. The rigid image registration was performed on Amira 3D with metrics of extended mutual information.

The image quality was evaluated using the most widely used metrics, the signal-to-noise ratio (PSNR) and the structure similarity index (SSIM) \cite{Zhao2019_TIP_MRISR_CSN,Zheng2021_CVPR_MRISR_SERAN,Pham2017_ISBI_MRISR_DenseNet,Chen2018_ISBI_MRISR_DenseNet,Chen2018_MICCAI_MRISR_GAN,Sui2020_MICCAI_MRISR_Gradient_Guidance,Feng2021_MICCAI_MRISR_Multi_Contrast}. The higher values of PSNR and SSIM represent better performance. Unlike regular image SRR studied in computer vision research, we didn't adopt any perceptual metrics since structural accuracy is more important than the pure perceptual effect in clinical diagnosis and image-guided treatments.

\section{Experiments and Results}

\subsection{Illustration of Degradation Adaptation}

As shown in Fig. \ref{degradation_illustration}, images with similar anatomical structures were selected from different participants in the source group (in the top row) and the target group (in the bottom row). $\textbf{Y}_{\textbf{s}}$ was the HR image from the source group and $\textbf{Y}_{\textbf{t}}$ was the SR image reconstructed using the LR image ($X_t$) from the target group. $X_{{s}\sim{t}}$ was the LR image downsampled from $\textbf{Y}_{\textbf{s}}$ by the degradation learning network of UDEAN. $X_t$ and $X_{{s}\sim{t}}$ had highly comparable image quality, this was achieved by the degradation adaptation mechanism of the network. $f_{s1}$ to $f_{s3}$ and $f_{t1}$ to $f_{t3}$ were the feature maps extracted from $\textbf{Y}_{\textbf{s}}$ and $X_t$, respectively. The patterns in the feature maps of the source group and the target group were highly consistent, revealing the effectiveness of the degradation adaptation in the latent feature space. With a certain consistency between the source group and the target group in both the image space and the latent feature space, UDEAN was able to reconstruct SR images from LR images from the target group with a quality comparable to that of HR images from the source domain. The degradation adaptation process can also be observed using t-SNE to show the distribution of extracted features in the feature space before and after degradation adaptation, as shown in Fig. \ref{DA_illustration}.

\subsection{Ablation Study of the UDEAN}

We evaluated our network in various configurations on the same SRR task for misaligned LR and HR datasets to find the best performance. The effect of DA in the image space and latent feature space was evaluated. Furthermore, we compared our network to the same backbone network trained in the supervised strategy with misaligned training data to reveal the effect of the degradation learning modules. As shown in Table \ref{tab1}, all the configurations of the UDEAN outperformed the supervised learning using the same generator by over 0.014/0.74 dB in SSIM/PSNR using misaligned LR and HR image from the HCP dataset with the scale factor of $2\times2\times2$. Among the configurations of the UDEAN, the performance was downgraded by 0.009/0.75 dB or 0.002/0.71 dB in SSIM/PSNR when DA in only latent feature space or image space was applied, respectively.

Furthermore, Fig. \ref{comparison_UDEAN_and_supervised_approaches} shows the visual effect of the reconstructed SR images and the error maps to the HR images. The SR images reconstructed by the supervised methods were highly blurry with or without rigid image registration. On the contrary, the quality of SR images reconstructed by the UDEAN was significantly improved even with DA only in latent feature space or image space. The errors of the reconstructed anatomical structures were further reduced when DA was applied in both domains in the training process.

\begin{table}[ht]
\centering
\caption{Comparison of model efficiencies for inference with the scale factor of $2\times2\times2$.}
\label{tab2}
\setlength{\tabcolsep}{6pt}
\begin{tabular}{c|c|c|c}
\hline
                            & \# Params (M) & GFlops & Inference Time (s) \\
\hline
ZSSR \cite{ZSSR}            & $2.457$ & $10.055$ & $2.26\pm0.04$ \\
DASR \cite{DASR}            & $2.457$ & $10.055$ & $2.33\pm0.03$ \\
PseudoSR \cite{Pseudo_SR}   & $4.765$ & $19.455$ & $5.27\pm0.07$ \\
Blind-SR \cite{BlindSR}     & $4.765$ & $19.455$ & $5.23\pm0.06$ \\
$\mathbf{UDEAN (ours)}$     & $2.457$ & $10.055$ & $2.30\pm0.03$ \\
\hline
\end{tabular}
\end{table}

\subsection{Comparison with the State-of-the-Art Unsupervised Methods}

\begin{table*}[ht]
\centering
\caption{Quantitative comparison with the state-of-the-art unsupervised networks for MRI SRR with misaligned and unpaired LR and HR images}
\label{tab3}
\setlength{\tabcolsep}{14pt}
\begin{tabular}{c|c|l|c|c c }
\hline
 Dataset & Scale Factor & Method & Misaligned or Unpaired & SSIM & PSNR \\
\hline
\hline
\multirow{20}*{HCP} & \multirow{10}*{$2\times2\times2$} & Tricubic & \multirow{2}*{N/A} & $0.8981\pm0.0106$ & $31.5862\pm1.8520$ \\
 & & ZSSR \cite{ZSSR} &  & $0.8994\pm0.0163$ & $33.0213\pm2.2054$ \\
\cline{3-6}
 & & DASR \cite{DASR} & \multirow{4}*{Unpaired} & $0.8931\pm0.0094$ & $32.3309\pm1.4671$ \\
 & & Pseudo SR \cite{Pseudo_SR} &  & $0.8931\pm0.0116$ & $31.0407\pm1.8816$\\
 & & Blind-SR \cite{BlindSR} &  & $0.9132\pm0.0100$ & $33.0695\pm1.4656$ \\
 & & $\mathbf{UDEAN (ours)}$ &  & $\mathbf{0.9242\pm0.0083}$ & $\mathbf{33.5159\pm1.7234}$ \\
 \cline{3-6}
 & & DASR \cite{DASR} & \multirow{4}*{Misaligned} & $0.8946\pm0.0091$ & $32.2544\pm1.4658$ \\
 & & Pseudo SR \cite{Pseudo_SR} &  & $0.8962\pm0.0086$ & $32.0267\pm1.5950$ \\
 & & Blind-SR \cite{BlindSR} &  & $0.9149\pm0.0091$ & $33.1597\pm1.5693$ \\
 & & $\mathbf{UDEAN (ours)}$ &  & $\mathbf{0.9247\pm0.0097}$ & $\mathbf{34.2729\pm1.8080}$ \\
\cline{2-6}
 & \multirow{10}*{$2\times2\times1$} & Tricubic & \multirow{2}*{N/A} & $0.9227\pm0.0088$ & $33.6545\pm1.8492$ \\
 & & ZSSR \cite{ZSSR} &  & $0.9412\pm0.0106$ & $\mathbf{36.0848\pm2.4098}$ \\
\cline{3-6}
 & & DASR \cite{DASR} & \multirow{4}*{Unpaired} & $0.9137\pm0.0084$ & $33.6732\pm1.4502$ \\
 & & Pseudo SR \cite{Pseudo_SR} &  & $0.9244\pm0.0076$ & $33.8122\pm1.6044$ \\
 & & Blind-SR \cite{BlindSR} &  & $0.9325\pm0.0081$ & $34.9527\pm1.7325$ \\
 & & $\mathbf{UDEAN (ours)}$ &  & $\mathbf{0.9371\pm0.0078}$ & $\mathbf{35.1688\pm1.8195}$ \\
\cline{3-6}
 & & DASR \cite{DASR} & \multirow{4}*{Misaligned} & $0.9192\pm0.0082$ & $33.9043\pm1.6386$ \\
 & & Pseudo SR \cite{Pseudo_SR} &  & $0.9228\pm0.0080$ & $34.0500\pm1.6694$ \\
 & & Blind-SR \cite{BlindSR} &  & $0.9395\pm0.0071$ & $35.0236\pm1.8130$ \\
 & & $\mathbf{UDEAN (ours)}$ &  & $\mathbf{0.9455\pm0.0072}$ & $\mathbf{35.5182\pm2.2839}$ \\
\hline
\hline
\multirow{20}*{BraTS} & \multirow{10}*{$2\times2\times2$} & Tricubic & \multirow{2}*{N/A} & $0.9073\pm0.0196$ & $31.2470\pm1.9187$ \\
 & & ZSSR \cite{ZSSR} &  & $0.9001\pm0.0464$ & $29.6156\pm3.1528$ \\
  \cline{3-6}
 & & DASR \cite{DASR} & \multirow{4}*{Unpaired} & $0.9334\pm0.0141$ & $30.5198\pm2.0642$ \\
 & & Pseudo SR \cite{Pseudo_SR} &  & $0.8978\pm0.0129$ & $26.4210\pm0.8741$ \\
 & & Blind-SR \cite{BlindSR} &  & $0.9434\pm0.0121$ & $31.2900\pm2.2618$ \\
 & & $\mathbf{UDEAN (ours)}$ &  & $\mathbf{0.9471\pm0.0096}$ & $\mathbf{31.4467\pm2.3570}$ \\
  \cline{3-6}
 & & DASR \cite{DASR} & \multirow{4}*{Misaligned} & $0.9377\pm0.0151$ & $31.1546\pm2.0495$ \\
 & & Pseudo SR \cite{Pseudo_SR} &  & $0.9378\pm0.0150$ & $31.0804\pm1.9391$ \\
 & & Blind-SR \cite{BlindSR} &  & $0.9360\pm0.0120$ & $31.2894\pm1.9017$ \\
 & & $\mathbf{UDEAN (ours)}$ &  & $\mathbf{0.9513\pm0.0133}$ & $\mathbf{32.0435\pm2.2796}$ \\
\cline{2-6} 
 & \multirow{10}*{$2\times2\times1$} & Tricubic & \multirow{2}*{N/A} & $0.9299\pm0.0151$ & $34.8846\pm2.2722$ \\
 & & ZSSR \cite{ZSSR} &  & $0.9514\pm0.0156$ & $33.2469\pm3.6422$ \\
  \cline{3-6}
 & & DASR \cite{DASR} & \multirow{4}*{Unpaired} & $0.9462\pm0.0149$ & $34.4395\pm1.8905$ \\
 & & Pseudo SR \cite{Pseudo_SR} &  & $0.9397\pm0.0150$ & $34.0382\pm1.9015$ \\
 & & Blind-SR \cite{BlindSR} &  & $0.9510\pm0.0156$ & $35.3235\pm2.0213$ \\
 & & $\mathbf{UDEAN (ours)}$ &  & $\mathbf{0.9551\pm0.0166}$ & $\mathbf{35.4973\pm2.2433}$ \\
  \cline{3-6}
 & & DASR \cite{DASR} & \multirow{4}*{Misaligned} & $0.9461\pm0.0167$ & $34.5953\pm2.0557$ \\
 & & Pseudo SR \cite{Pseudo_SR} &  & $0.9463\pm0.0162$ & $34.5560\pm1.9037$ \\
 & & Blind-SR \cite{BlindSR} &  & $0.9558\pm0.0158$ & $36.1029\pm2.2270$ \\
 & & $\mathbf{UDEAN (ours)}$ &  & $\mathbf{0.9629\pm0.0163}$ & $\mathbf{36.7631\pm2.3787}$ \\
\hline
\end{tabular}
\end{table*}

\begin{figure*}
\centering
\includegraphics[width=\textwidth]{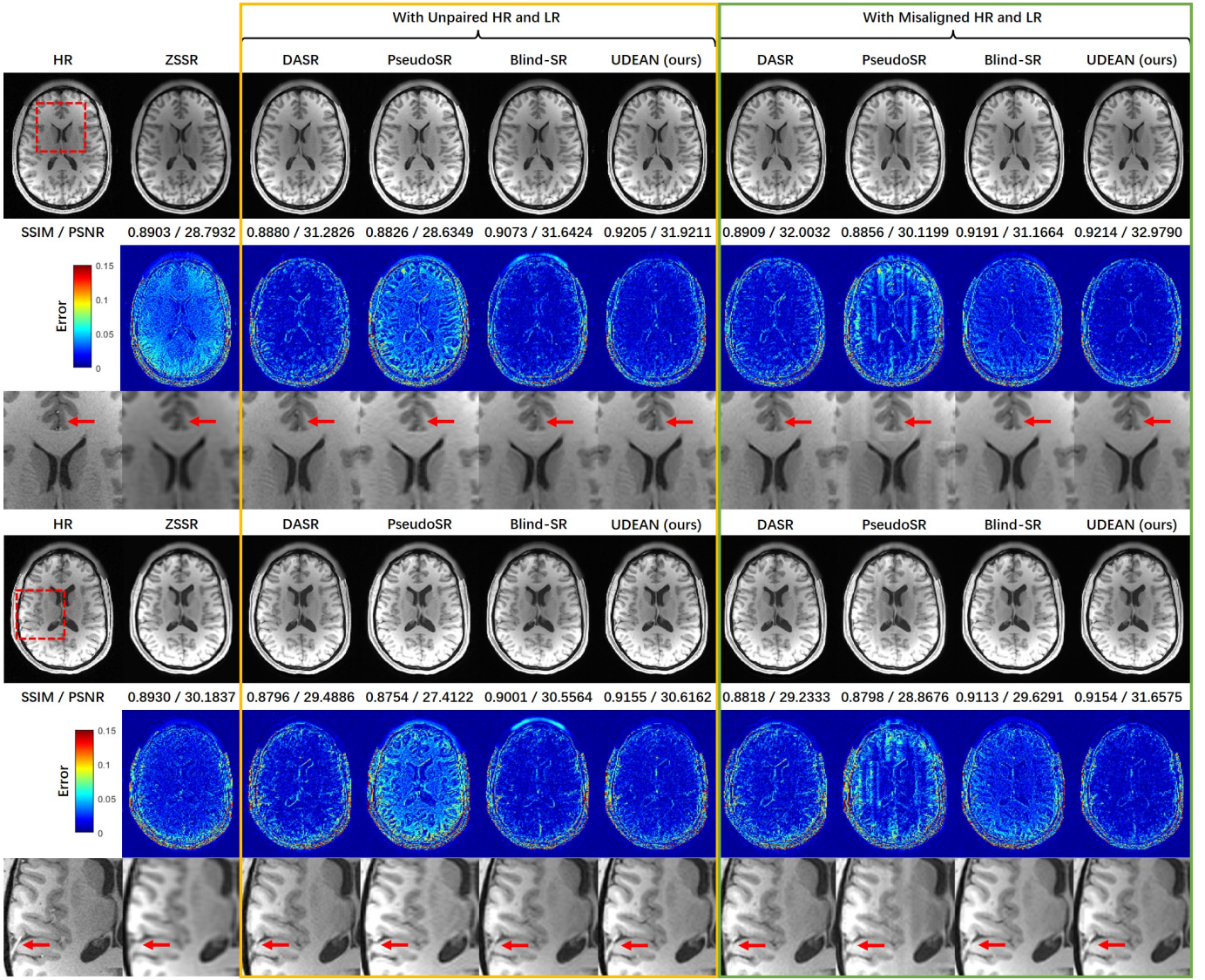}
\caption{Qualitative comparison of the UDEAN with the state-of-the-art unsupervised networks in visual effect and error maps in the axial plane with the HCP dataset and the scale factor of $2\times2\times2$. The arrows point to the small structures that UDEAN reconstructed with higher accuracy than the other networks.} 
\label{Fig_Comparison_other_unsupervised_methods}
\end{figure*}

\begin{figure*}
\centering
\includegraphics[width=\textwidth]{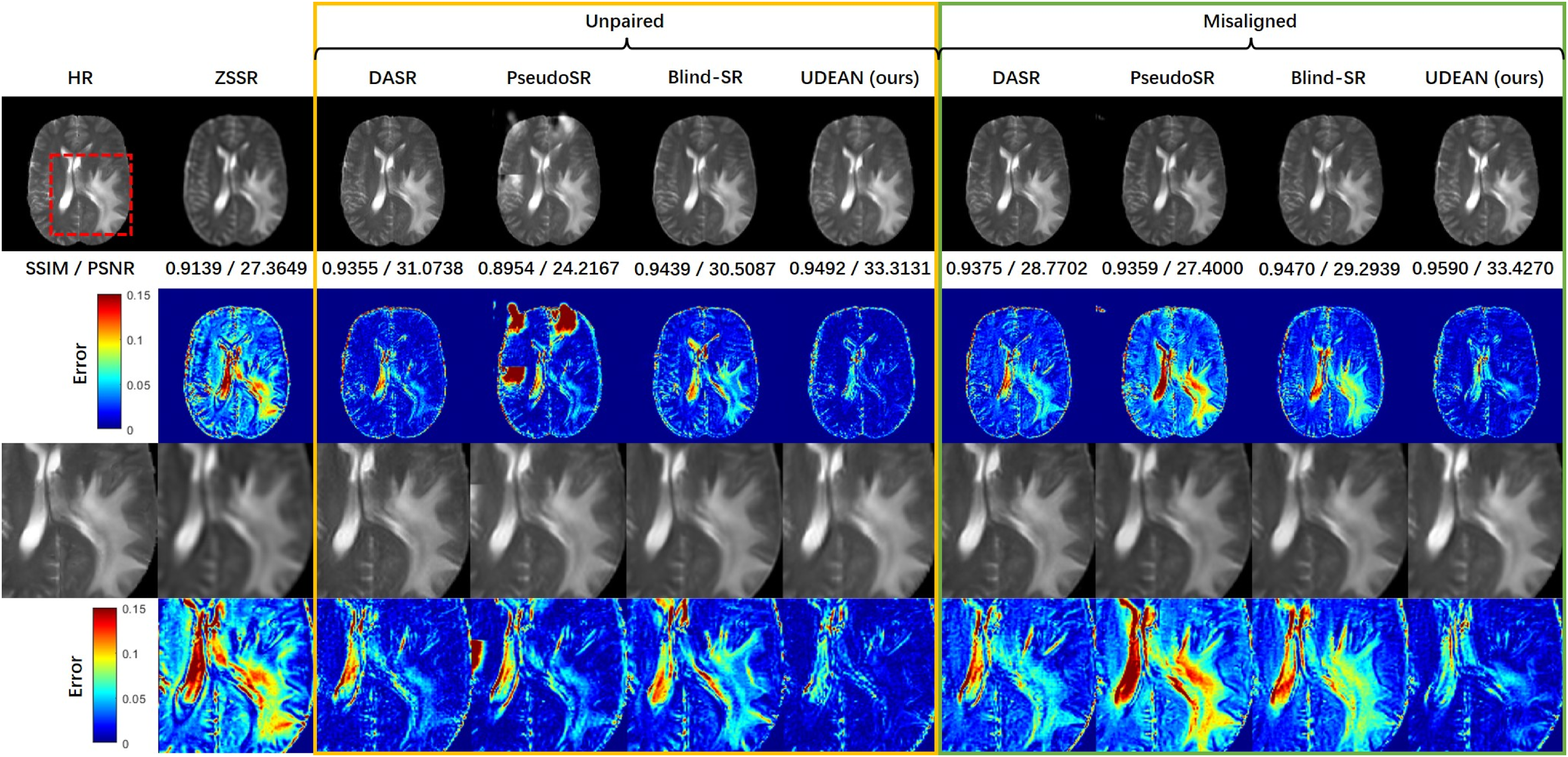}
\caption{Qualitative comparison of the UDEAN with the state-of-the-art unsupervised networks in visual effect and error maps in the axial plane with the BraTS dataset and the scale factor of $2\times2\times2$. UDEAN achieved the lowest error in the region of the lesion. }
\label{Fig_sota_comparison_brats}
\end{figure*}

We implemented other state-of-the-art unsupervised networks, including ZSSR \cite{ZSSR},  DASR \cite{DASR}, PseudoSR \cite{Pseudo_SR}, and Blind-SR \cite{BlindSR}, as the baseline. All these networks used the same backbone as the UDEAN for both generators and discriminators and trained with identical training setups, including training datasets and hyperparameters, for fair comparisons. For inference, UDEAN and DASR used a single upsampling network with the same structure. As a result, the inference process for UDEAN and DASR consumed identical computational resources, including the same number of network parameters, number of operations, GPU consumption, and inference time. Blind-SR and PseudoSR used another network for style transfer in addition to the upsampling network, thus consuming approximately twice as many computation resources as the other two. More detailed numerical comparisons on number of parameters, FLOPS, and inference time, are shown in Table \ref{tab2}. The numerical results in Table \ref{tab3} reveal that the UDEAN outperformed all the above-mentioned networks for MRI SRR from both misaligned and unpaired training data, which are from two public datasets with different contrasts and two scale factors. Specifically, for the HCP dataset and both scale factors, UDEAN outperformed the other networks in the SSIM and PSNR values for both misaligned and unpaired training data, only except the PSNR value of ZSSR with the scale factor of $2\times2\times1$. For the BraTS dataset, the UDEAN also outperformed all other networks with the highest SSIM and PSNR values in most experiments.

As for the qualitative comparison shown in Fig. \ref{Fig_Comparison_other_unsupervised_methods} for the SRR of the HCP dataset with a scale factor of $2\times2\times2$, ZSSR generated highly blurry images, although its metrics values are higher than DASR and PseudoSR in some cases. Among all the images, the ones reconstructed by UDEAN achieved the lowest errors as shown in the error maps, and the best accuracy in small anatomical structures pointed out by the red arrows. Also shown in Fig. \ref{Fig_sota_comparison_brats} for the SRR with the BraTS dataset and the scale factor of $2\times2\times2$, UDEAN achieved better accuracy than the other networks in the structures of lesions, which may improve the accuracy of the assessment of the lesions and the therapy based on the reconstructed SR images.

\section{Discussion}

\subsection{Practicality in Clinical Settings}
In real clinical settings, perfectly paired and aligned authentic LR and HR images are difficult to acquire due to patient movements. Besides, in-vivo images are mostly acquired with either LR or HR images in clinical settings because of high time consumption and cost. Lack of paired and aligned LR and HR images raises challenges in implementing supervised SRR methods. Our experimental results revealed that supervised methods could not provide acceptable image quality if the LR and HR images in the training datasets were not perfectly aligned. Even with rigid image registration, the inference results were still unsatisfactory if there was a tiny geometric deformation between the LR and HR images. Besides, considering the change in object size in our experiments was between 0.7\% and 0.9\%, which was far beyond the accuracy of 90\% DSC achieved by the latest deep learning-based deformable image registration \cite{Segmentation_accuracy}, the deformable image registration algorithms are not helpful in solving such tiny misalignments of soft tissues. Therefore, previous studies adopted synthetic LR images generated from HR images with certain downsampling algorithms as the input of the network \cite{Zhao2019_TIP_MRISR_CSN,Zheng2021_CVPR_MRISR_SERAN,Pham2017_ISBI_MRISR_DenseNet,Chen2018_ISBI_MRISR_DenseNet,Chen2018_MICCAI_MRISR_GAN,Sui2020_MICCAI_MRISR_Gradient_Guidance,Feng2021_MICCAI_MRISR_Multi_Contrast}. However, the degradation shift between the artificial downsampling and the real degradation from authentic HR to LR images downgrades the performance of the trained networks when they are used for SRR from authentic LR images \cite{DA_MRISR_ETHZ}. Therefore, unsupervised networks can be a possible solution to these problems. 

Previously proposed unsupervised networks were designed for 2D real-world images \cite{ZSSR,Cyclegan,DASR,Pseudo_SR}. However, medical images are mostly in 3D form, containing through-slice information. Therefore, we adopted the TS-RCAN \cite{TS-RCAN}, which is a low-cost network for 3D SRR with enhanced performance, as the backbone of the UDEAN to process 3D medical images. And compared to the other unsupervised method,  UDEAN learned the degradation representation in both the image space and the latent feature space, thus achieving better performance.

\subsection{Effect of Degradation Representation Learning}
The degradation representation learning of the UDEAN was a specific type of domain adaptation, which transferred HR images of the source domain to the LR images of the target domain. As illustrated in the result section, UDEAN was able to downsample the HR image from the source group to the LR image, which had comparable quality to the target group. Consistency in the latent feature space between the source group and the target group further helped the network with improved accuracy. The ablation study also revealed the effects of domain adaptations, which guided the network to learn the degradation representation in both the image and the latent feature space. The experimental results showed the advantage of UDEAN over the supervised learning approach in misaligned LR and HR images. When the LR and HR images of the training datasets were not perfectly aligned, the supervised methods failed to provide acceptable image quality. Besides, the ablation study also revealed the effect of degradation representation learning in both image and latent feature space. The network could reconstruct the SR images without either $Loss_{da}$ or $Loss_{fa}$. However, the absence of $Loss_{da}$ or $Loss_{fa}$ downgraded the network's performance, showing the benefit of degradation representation learning in both the image space and the latent feature space.

\subsection{Comparison of Different Domain Transfer Strategies}

Regarding the other state-of-the-art unsupervised networks, the ZSSR, as a network without domain transfer, learned the degradation representation between the LR images and the lower-resolution images, which were downsampled from the LR images, and used the learned degradation representation to reconstruct the SR images from the LR images \cite{ZSSR}. However, there was a large gap between the learned degradation and the degradation from HR images to the LR images, thus leading to downgraded performance, particularly with large-scale factors. Although it achieved high metrics values in a few cases, the reconstructed SR images were highly blurry as shown in our results, and cannot meet the quality required in clinics.

Most of the unsupervised networks with domain transfer are derived from CycleGAN. As one of the latest unsupervised SR approaches, the DASR learns the degradation representation by downsampling the HR images of the source domain to LR images in the target domain, and domain distance maps in the image space are generated simultaneously, which are subsequently used to reconstruct SR images \cite{DASR}. However, this learning strategy was not end-to-end, so the early stop when the downsampling network (DSN) reached the best performance in the validation dataset was not applicable. The DSN could be overfitted to the training dataset, making the whole network difficult to train and optimize, and the learned degradation representation hardly fits the reconstruction procedure. As a result, the inaccuracies in the DASR's reconstructed SR MRI images are still massively visible. 

PseudoSR is another most recent approach to an unsupervised SRR network. It transfers the LR image of the target domain to the source domain, where the LR images are downgraded from HR images with a predefined algorithm (normally bicubic or Gaussian blurring). Then the SR images are reconstructed from downsampled LR images \cite{Pseudo_SR}. The experimental results showed that the PseudoSR could achieve neither high metrics values nor satisfactory qualitative image quality. Despite the adaptation of latent feature space, another main difference between the UDEAN and the PseudoSR was that the UDEAN transfers the source domain HR images to the target domain and learns the mapping of the LR images from the target domain to the HR images in the source domain. On the contrary, the PseudoSR transfers the target domain LR images to the source domain and learns the mapping of the LR images from the source domain to the HR images in the source domain. The ablation study results showed the advantage of the former scheme even with the absence of feature space adaptation. And with the latter approach, more detailed structures were missing in the reconstructed SR images.

Blind-SR is another variation of CycleGAN with simple modifications proposed after UDEAN\footnote{https://arxiv.org/abs/2205.06891v1, released on 13 May 2022.}. It shares a similar scheme with the PseudoSR, which transfers the LR image of the target domain to the source domain, whose images are downsampled with bicubic or nearest algorithms. Therefore, it faced a similar problem to PseudoSR, with which much detailed structural information was lost. In addition, the Blind-SR was not trained in an end-to-end fashion either. Its two components were trained separately, making it face similar difficulties in optimizing the network like DASR. Moreover, Blind-SR showcases performance degradation from UDEAN in both SSIM and PNSR, as well as low quality of MRI SRR in visual effect as shown in Fig.\ref{Fig_Comparison_other_unsupervised_methods} and Fig.\ref{Fig_sota_comparison_brats}. The aforementioned drawbacks make Blind-SR impractical in clinical treatment.

Besides the differences in the domain transfer strategies, UDEAN and DASR utilized a single upsampling network for the inference, whereas PseudoSR and Blind-SR used another style transfer network in addition to the upsampling network. As a result, the computation resources consumed by UDEAN were identical to DASR and approximately 50\% less than PseudoSR and Blind-SR.

\subsection{Limitations and Future Works}
Despite the promising results of UDEAN, we acknowledge certain limitations in this study. We only tested the UDEAN on brain MRI images, where the geometric deformation is normally mild in real clinics, and the LR images were synthetically generated from HR images. Abdominal and musculoskeletal imaging normally suffer from more severe geometric deformation of soft tissues due to irresistible patient movements. Therefore, the UDEAN can be the best solution in these scenarios and will be investigated in our future study. Besides, public datasets containing authentic LR and HR images of abdominal and musculoskeletal imaging are unavailable. Therefore, the misaligned or unpaired LR and HR images still need to be artificially generated. This study focused on investigating the potential of SRR using networks trained by misaligned or unpaired LR and HR images, the datasets we used in our experiments were still applicable to our experiments. Our future study will investigate the performance of UDEAN trained with authentic LR and HR images. 

\section{Conclusion}

In this manuscript, we propose the UDEAN as an unsupervised degradation adaptation network that adaptively learns the degradation representation between misaligned or unpaired LR and HR MRI data in both image and latent feature spaces. UDEAN applies SRR in an end-to-end fashion, making the network easy to train and optimize. The DA mechanism adopted by the UDEAN also shows advantages over previously used domain transfer methods, thus minimizing errors in the reconstructed SR images. Our experimental results showed that the UDEAN alleviated the problem of lacking paired authentic LR and HR images, achieved enhanced image quality for SRR, and outperformed the other state-of-the-art networks. Therefore, UDEAN is a promising solution for SRR in clinical settings when perfectly aligned LR and HR image pairs are unavailable. Despite the significant advancements in medical imaging, there is still a shortage of large-scale LR and HR MRI images measured from patients in the clinical environment. As a result, all current state-of-the-art methods, including ours, have been trained using authentic HR images and artificially generated LR images from authentic HR images. However, we plan to incorporate large-scale clinical LR images into our future work to enhance our studies.

\begin{IEEEbiography}[{\includegraphics[width=1in,height=1.25in,clip,keepaspectratio]{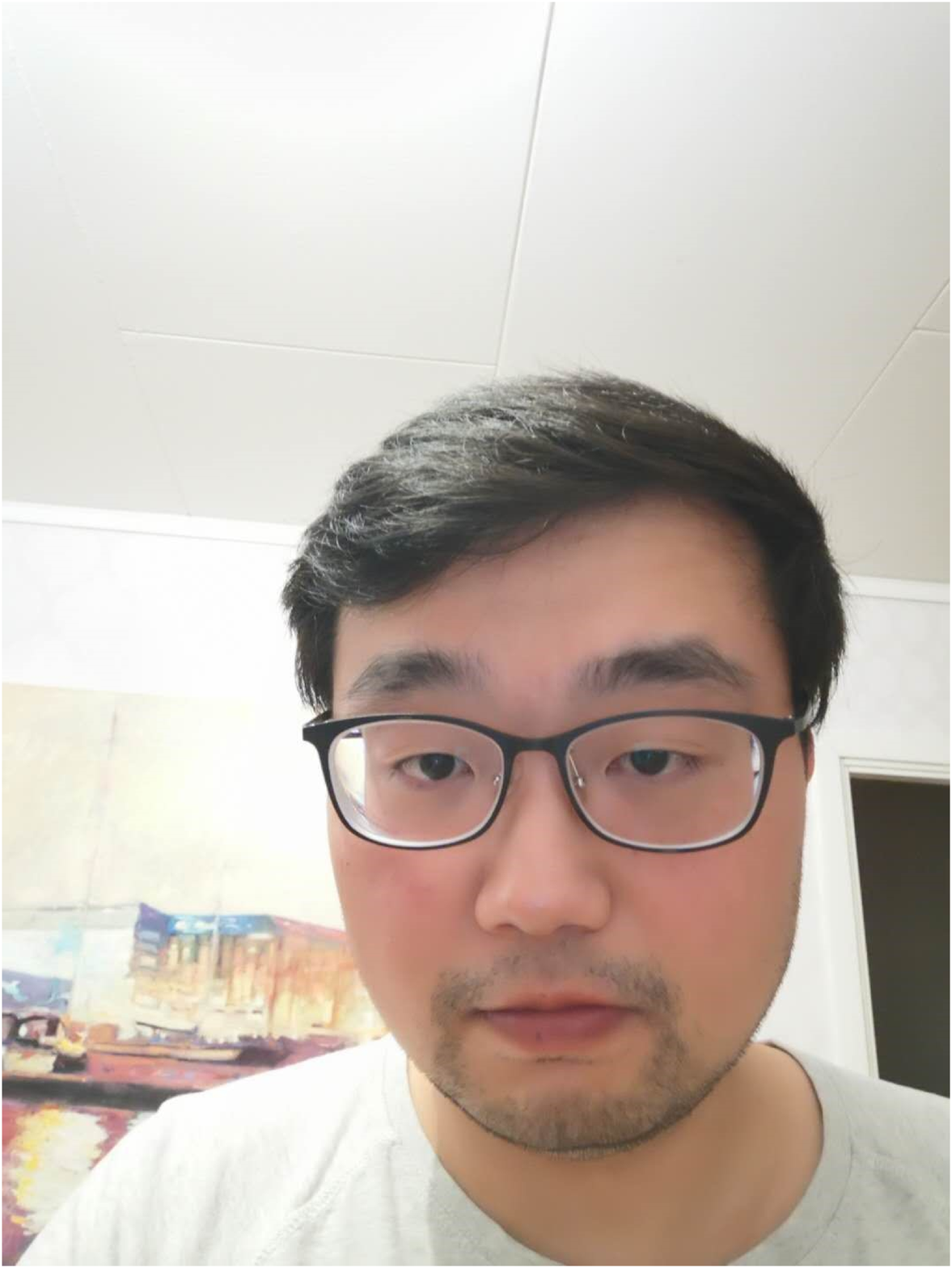}}]{Jianan Liu} received his B.Eng. degree in Electronics and Information Engineering from Huazhong University of Science and Technology, Wuhan, China, in 2007. He received his M.Eng. degree in Telecommunication Engineering from the University of Melbourne, Australia, and his M.Sc. degree in Communication Systems from Lund University, Sweden, in 2009 and 2012, respectively.
Jianan has over ten years of experience in software and algorithm design and development. He has held senior R\&D roles in the AI consulting, automotive, and telecommunication industries.
His research interests include applying statistical signal processing and deep learning for medical image processing, wireless communications, IoT networks, indoor sensing, and outdoor perception using a variety of sensor modalities like radar, camera, LiDAR, WiFi, etc.
\end{IEEEbiography}

\begin{IEEEbiography}[{\includegraphics[width=1in,height=1.25in,clip,keepaspectratio]{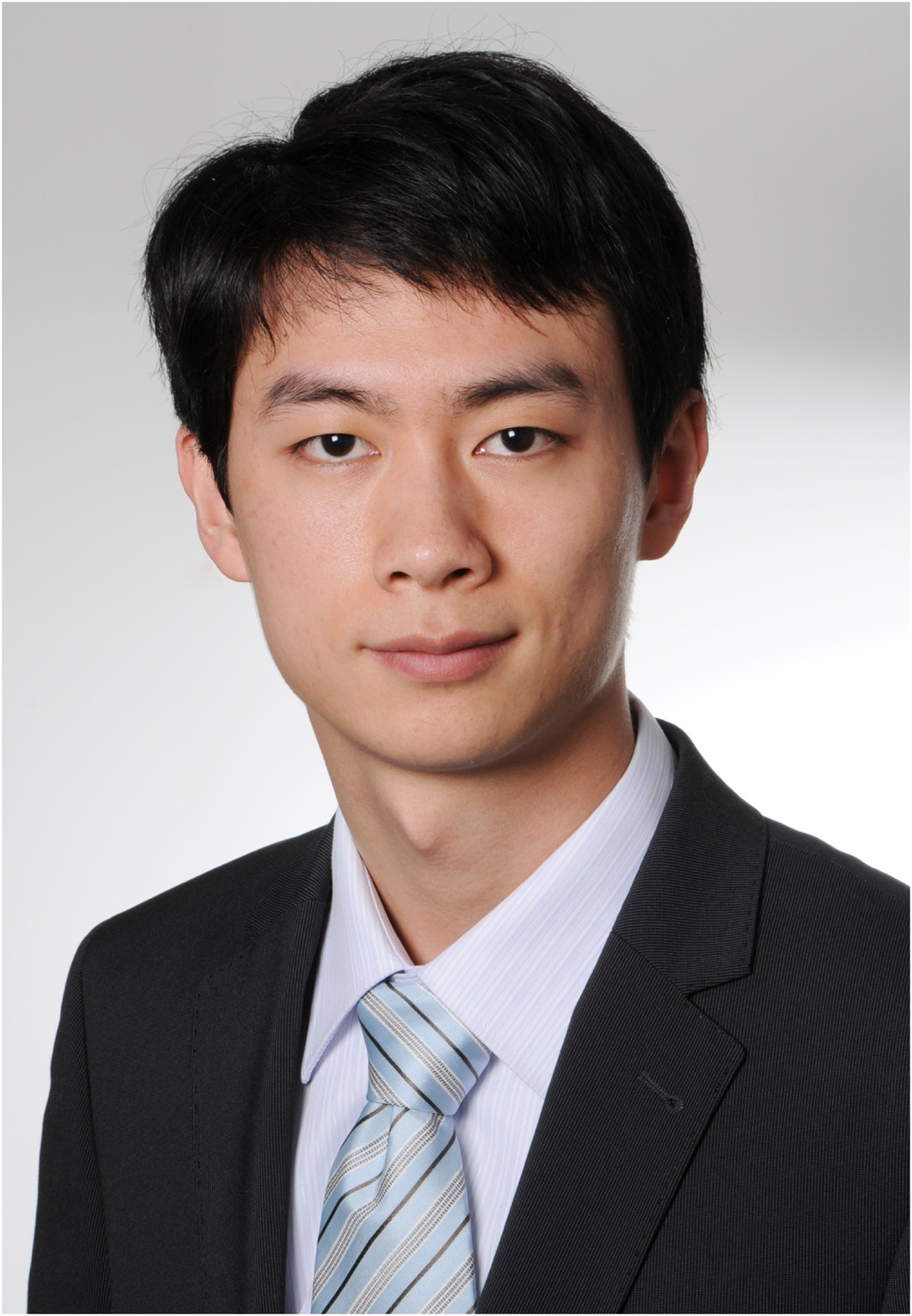}}]{Hao Li} received his B.Eng. degree in Electronics and Information Engineering from Huazhong University of Science and Technology, Wuhan, China, in 2007. He received his M.Sc. degree in Medical System Engineering from the University Magdeburg, Germany, in 2012. He is a research associate at the University Hospital Heidelberg, Germany. He has over ten years of experience in MRI pulse sequence design and medical image processing. He had held senior R\&D roles in the healthcare industry and clinical research. His research interests include MRI pulse sequence design, conventional and deep learning-based medical image processing, etc.
\end{IEEEbiography}

\begin{IEEEbiography}[{\includegraphics[width=1in,height=1.25in,clip,keepaspectratio]{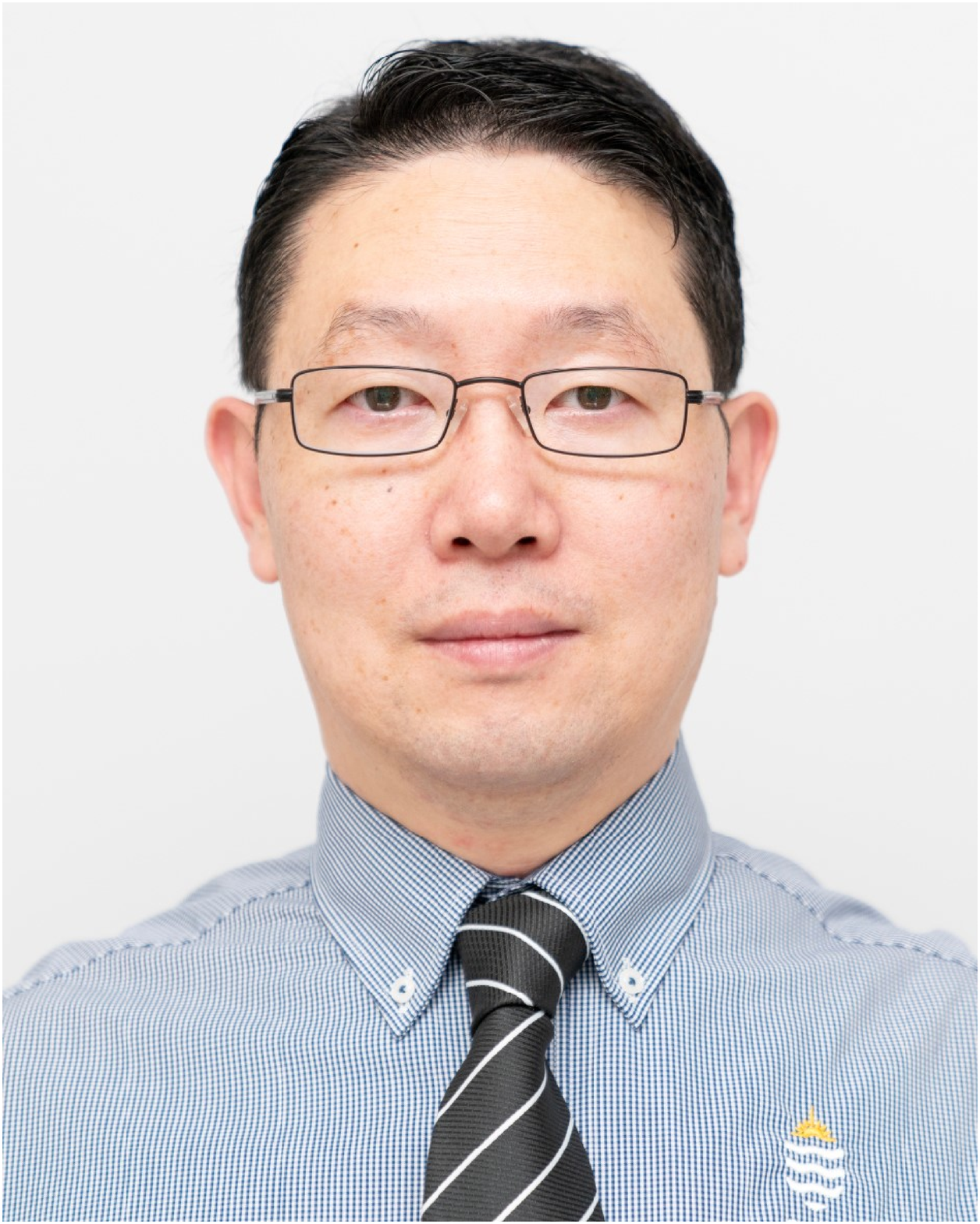}}]{Tao Huang}
(Senior Member, IEEE) holds a Ph.D. in Electrical Engineering from The University of New South Wales, Sydney, Australia. He also holds an M.Eng. in Sensor System Signal Processing from The University of Adelaide, Adelaide, Australia, and a B.Eng. in Electronics and Information Engineering from Huazhong University of Science and Technology, Wuhan, China.
Currently, Dr. Huang is a Senior Lecturer at James Cook University, Cairns, Australia, and serves as the course coordinator for the Master of Engineering (Professional). He was an Endeavour Australia Cheung Kong Research Fellow, a visiting scholar at The Chinese University of Hong Kong, a research associate at the University of New South Wales, and a postdoctoral research fellow at James Cook University. Before academia, Dr. Huang worked in the industry and held positions such as senior software engineer, senior data scientist, project team lead, and technical lead.
Dr. Huang has received the Australian Postgraduate Award, the Engineering Research Award at The University of New South Wales, the Best Paper Award from the IEEE Wireless Communications and Networking Conference, the IEEE Outstanding Leadership Award, and the Citation for Outstanding Contribution to Student Learning at James Cook University. In addition, he is a co-inventor of an international patent on MIMO systems.
Dr. Huang is a member of the IEEE Communications Society, IEEE Vehicular Technology Society, IEEE Computational Intelligence Society, and IEEE Industrial Electronics Society. He serves as Vice-Chair of the IEEE Northern Australia Section and local MTT-S/COM Chapter Chair. He also serves in various capacities at international conferences as TPC chair/vice chair, program vice chair, symposium chair, local chair, and TPC member. Dr. Huang is an Associate Editor of the IEEE Open Journal of Communications Society, IEEE Access, and IET Communications. He is also a Topical Advisory Panel Member and Guest Editor of MDPI Electronics. His research interests include electronics systems, deep learning, intelligent sensing, computer vision, pattern recognition, wireless communications, and IoT security.

\end{IEEEbiography}

\begin{IEEEbiography}[{\includegraphics[width=1in,height=1.25in,clip,keepaspectratio]{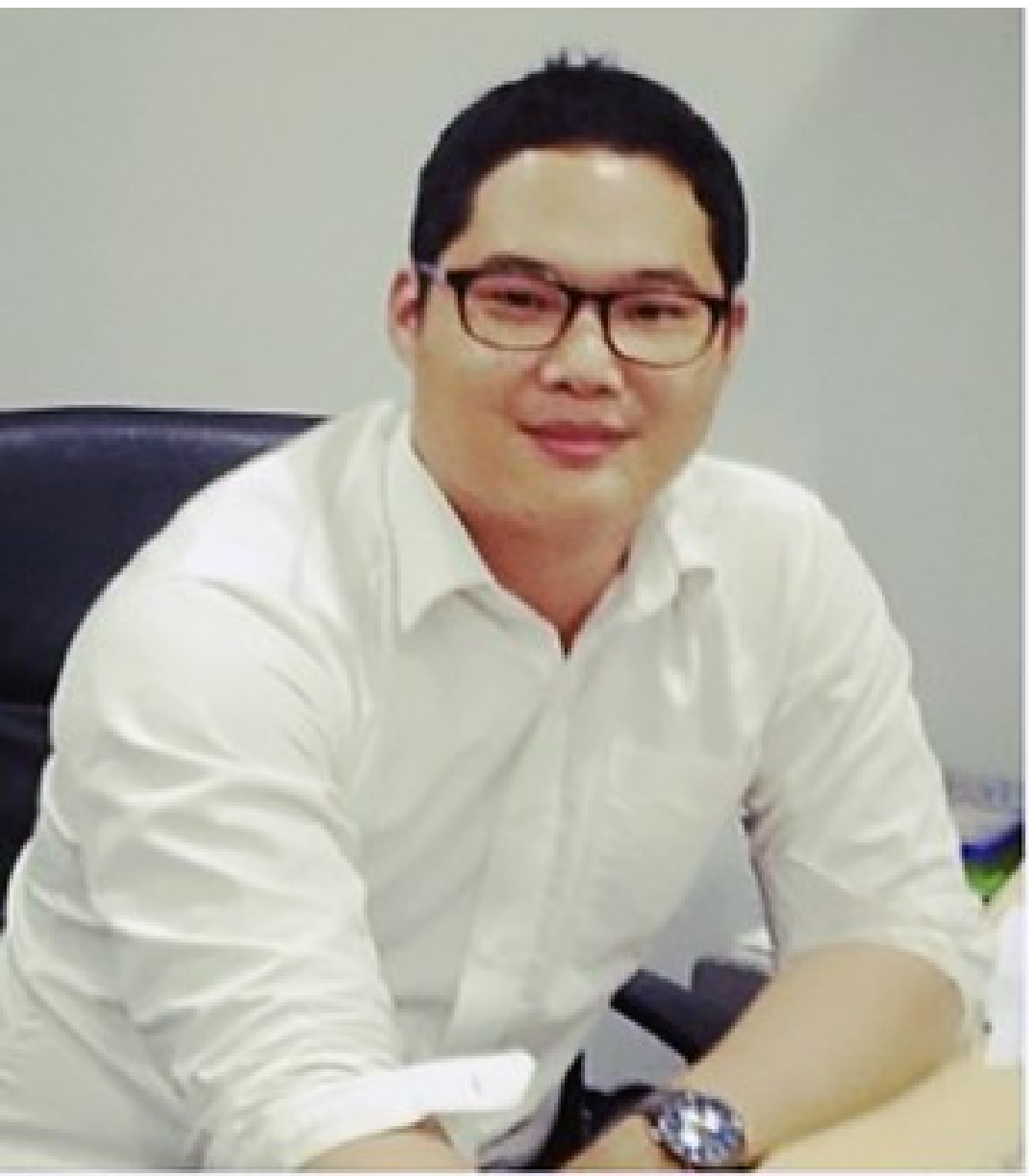}}]{Euijoon Ahn} is a Lecturer at the College of Science and Engineering, James Cook University. Prior to this, He was a postdoctoral research fellow at the Biomedical Multimedia Information Technology (BMIT) group, at the School of Computer Science, The University of Sydney. He obtained his Ph.D. degree in Computer Science (medical image analysis) from The University of Sydney in 2020. He received B. IT degree from The University of Newcastle, Australia, 2009, and M. IT (2014) and MPhil (2016) degree from The University of Sydney. His research focus is on the development of Machine Learning and Deep Learning, Computer Vision, and more specifically, unsupervised / self-supervised deep learning models for biomedical image analysis, for improving image segmentation, retrieval, quantification, and classification without relying on labeled data. He also works at the coalface of translational health technology research, e.g., health data analytics and telehealth. He has produced top-tier publications in the areas of computer vision and medical image computing, including papers in IEEE T-MI, T-BME, JBHI, MedIA, PR, CVPR, AAAI and MICCAI. 
\end{IEEEbiography}

\begin{IEEEbiography}[{\includegraphics[width=1in,height=1.25in,clip,keepaspectratio]{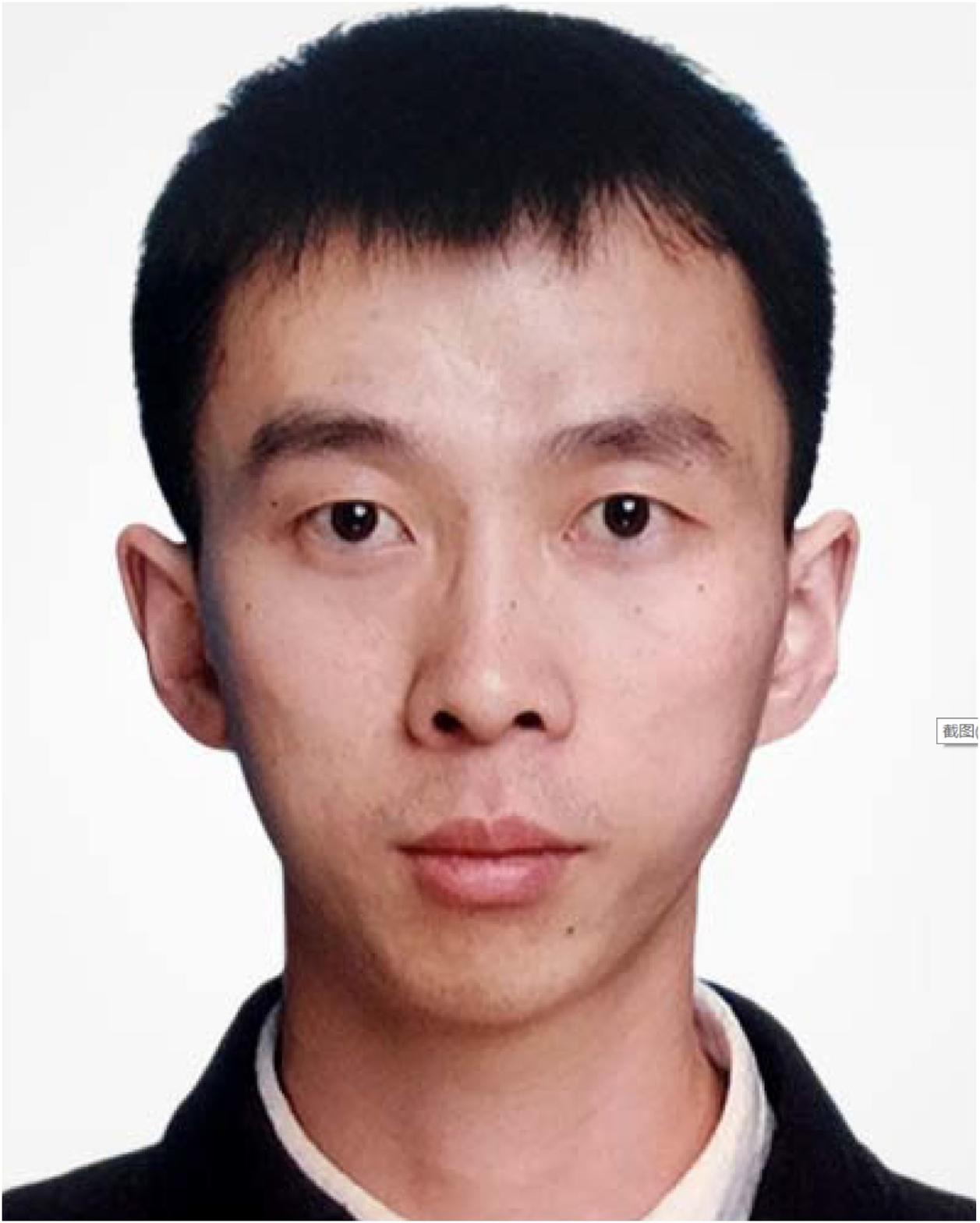}}]{Kang Han} received the B.Eng. degree in electronic
science and technology from Anhui Polytechnic University, Wuhu, China, in 2014, the M.Eng. degree in circuits and system from Shanghai University, Shanghai, China, in 2017, and the Ph.D. degree from James Cook University, Cairns, Australia, in 2023. He is currently a post-doctoral researcher at La Trobe University, Australia. His research areas include light field image processing, computer vision, and machine learning. In 2017, he won the Certificate of Science and Technology Achievement of Jiangxi Province, China.
\end{IEEEbiography}

\begin{IEEEbiography}[{\includegraphics[width=1in,height=1.25in,clip,keepaspectratio]{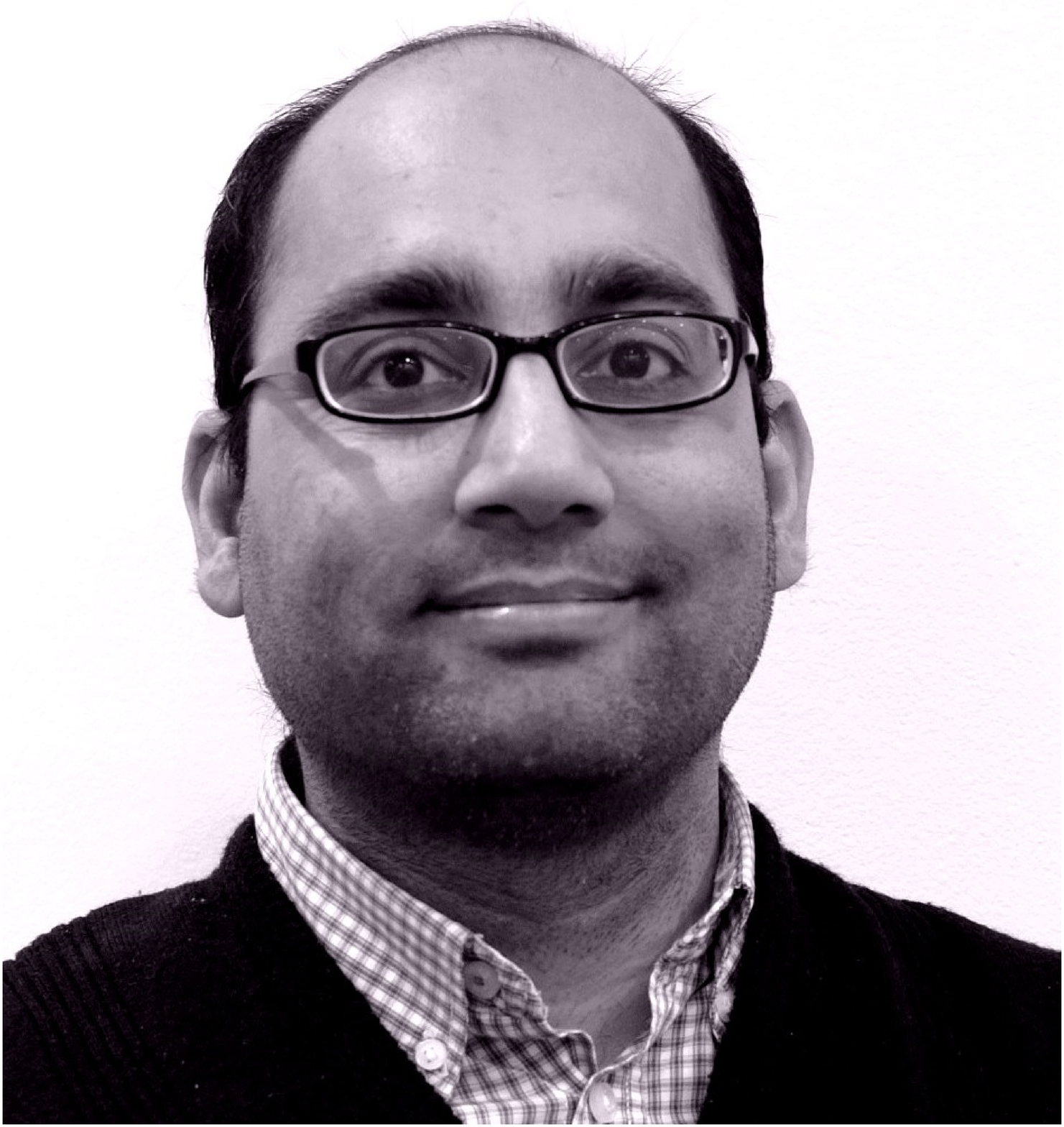}}]{Adeel Razi} is an Associate Professor at the Turner Institute for Brain and Mental Health, in the School of Psychological Sciences, Monash University, Australia. He joined Monash, after finishing his postdoctoral studies (2012-2018) at the Wellcome Centre for Human Neuroimaging, UCL, UK. His research is cross-disciplinary, combining engineering, physics, and machine-learning approaches, to model complex, multi-scale, network dynamics of brain structure and function using neuroimaging. He is currently an NHMRC Investigator (Emerging Leadership, 2021-2025), CIFAR Azrieli Global Scholar (2021-2023) in their Brain, Mind, and Consciousness Program, and an ARC DECRA Fellow (2018-2021). He received a B.E. degree in Electrical Engineering (with a Gold Medal) from the N.E.D. University of Engineering \& Technology in Pakistan, the M.Sc. degree in Communications Engineering from the University of Technology Aachen (RWTH), Germany, and the Ph.D. degree in Electrical Engineering from the University of New South Wales, Australia in 2012.
\end{IEEEbiography}

\begin{IEEEbiography}[{\includegraphics[width=1in,height=1.25in,clip,keepaspectratio]{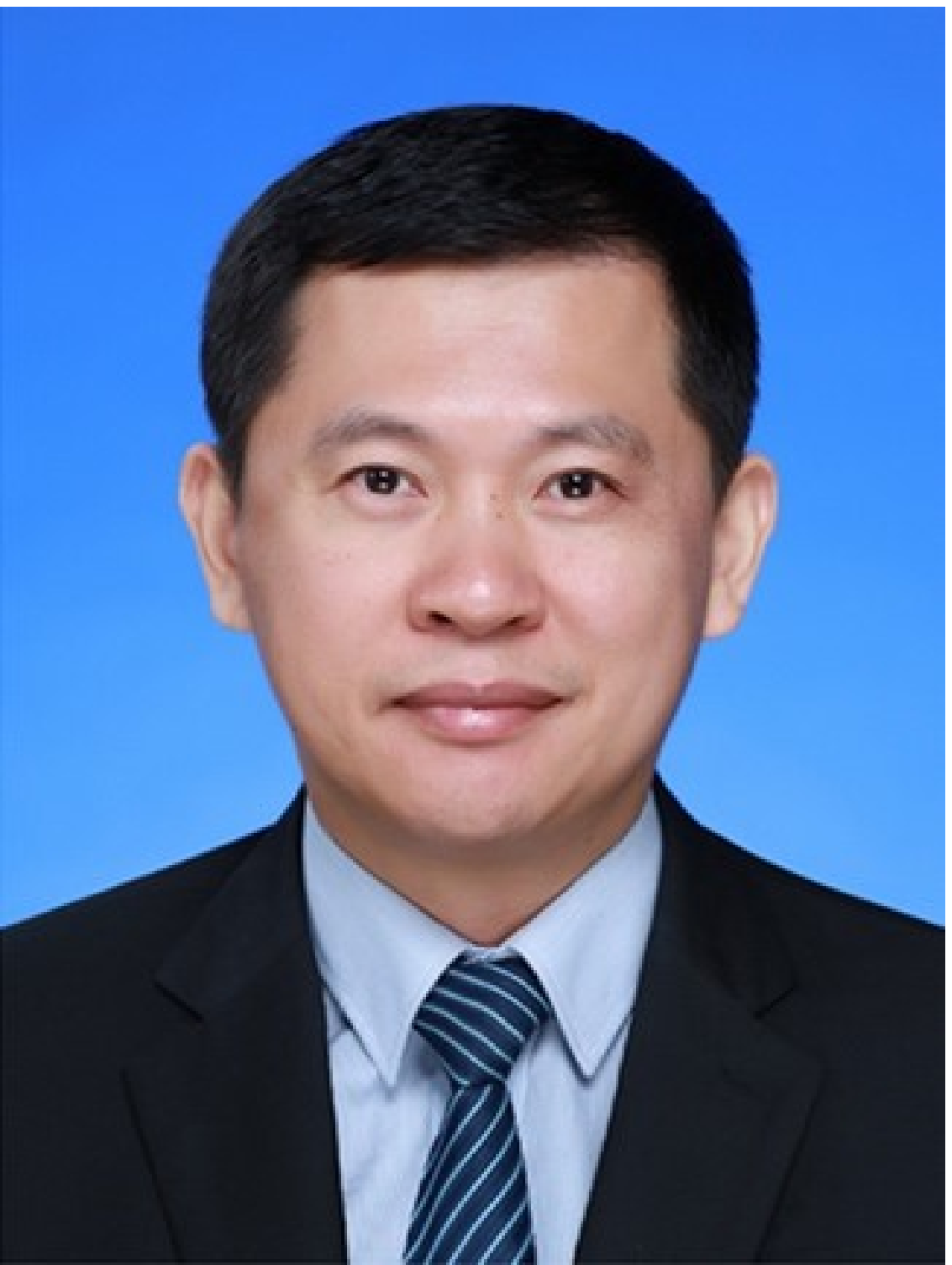}}]{Wei Xiang} (S’00–M’04–SM’10) Professor Wei Xiang is Cisco Research Chair of AI and IoT and Director of the Cisco-La Trobe Centre for AI and IoT at La Trobe University. Previously, he was Foundation Chair and Head of the Discipline of IoT Engineering at James Cook University, Cairns, Australia. Due to his instrumental leadership in establishing Australia’s first accredited Internet of Things Engineering degree program, he was inducted into Pearcy Foundation’s Hall of Fame in October 2018. He is a TEDx speaker and an elected Fellow of the IET in the UK and Engineers Australia. He received the TNQ Innovation Award in 2016, and Pearcey Entrepreneurship Award in 2017, and Engineers Australia Cairns Engineer of the Year in 2017. He was a co-recipient of four Best Paper Awards at WiSATS’2019, WCSP’2015, IEEE WCNC’2011, and ICWMC’2009. He has been awarded several prestigious fellowship titles. He was named a Queensland International Fellow (2010-2011) by the Queensland Government of Australia, an Endeavour Research Fellow (2012-2013) by the Commonwealth Government of Australia, a Smart Futures Fellow (2012-2015) by the Queensland Government of Australia, and a JSPS Invitational Fellow jointly by the Australian Academy of Science and Japanese Society for Promotion of Science (2014-2015). He was the Vice Chair of the IEEE Northern Australia Section from 2016-2020. He was an Editor for IEEE Communications Letters (2015-2017) and is currently an Associate Editor for IEEE Communications Surveys \& Tutorials, IEEE Internet of Things Journal, IEEE Access, and Nature Journal of Scientific Reports. He has published over 300 peer-reviewed papers including 3 books and 220 journal articles. He has severed in a large number of international conferences in the capacity of General Co-Chair, TPC Co-Chair, Symposium Chair, etc. His research interest includes the Internet of Things, wireless communications, machine learning for IoT data analytics, and computer vision.
\end{IEEEbiography}

\begin{IEEEbiography}[{\includegraphics[width=1in,height=1.25in,clip,keepaspectratio]{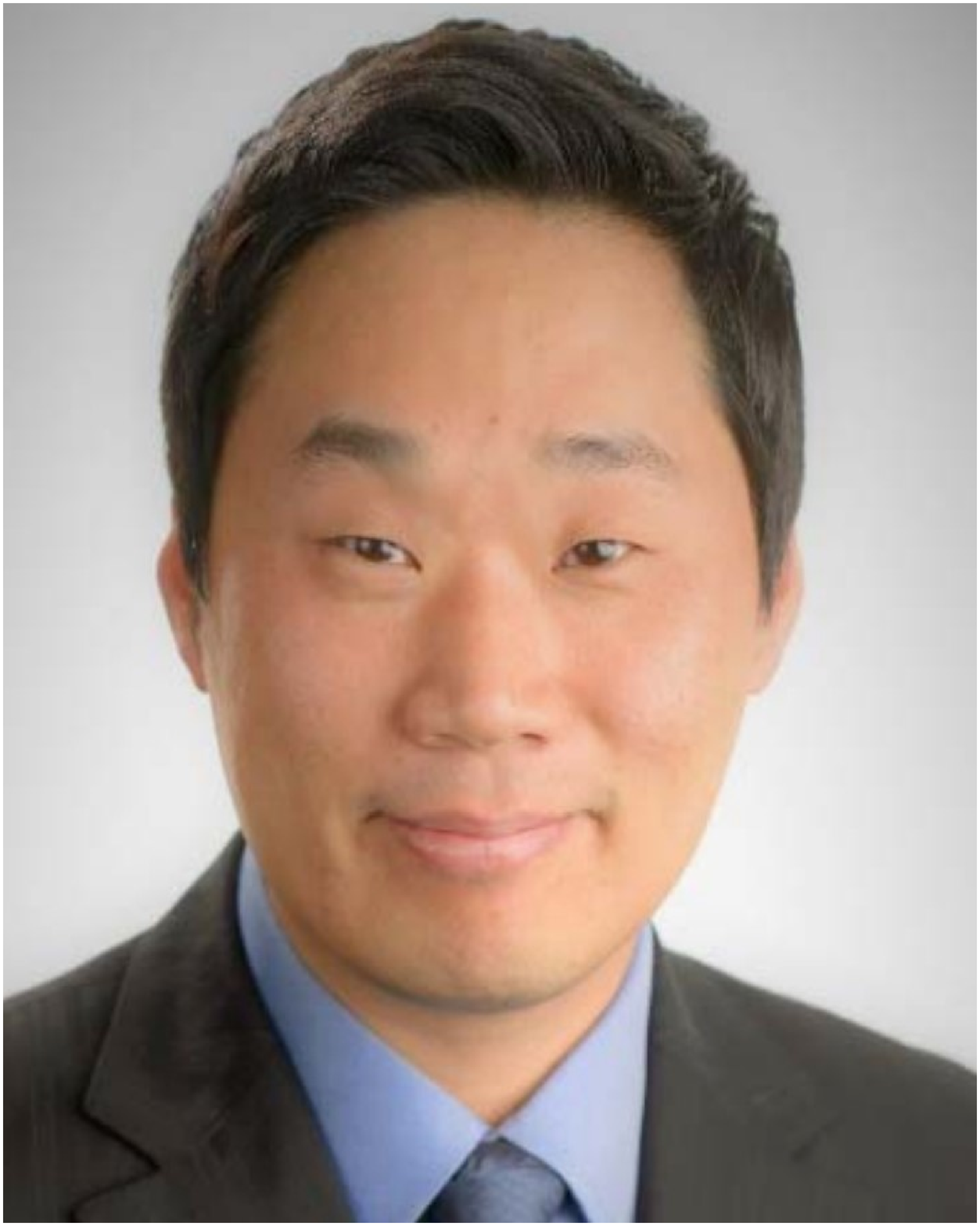}}]{Jinman Kim}(Member, IEEE) received the Ph.D. degree in computer science from The University of Sydney, Ultimo, NSW, Australia, in 2006. He was an ARC Post-Doctoral Research Fellow with The University of Sydney and then a Marie Curie Senior Research Fellow with the University of Geneva, Geneva, Switzerland. In 2013, he joined The University of Sydney as a Faculty Member. His research is in the application of machine learning for biomedical image analysis and visualization. His focus is on multimodal data processing and includes image omics, multimodal data processing, and image data correlation to other health data. He has produced a number of publications in this field and received multiple competitive grants and scientific recognitions. He has actively focused on research translation, where he has worked closely with clinical partners to take his research into clinical practice. He is the Research Director of the Nepean Telehealth Technology Centre (NTTC) at the Nepean hospital, responsible for translational telehealth and digital hospital research. Some of his research has been developed into clinical software that is being used/trialed at multiple hospitals. Dr. Kim is actively involved in his research communities, where he is an Associate Editor of TVCJ and the Vice President of the Computer Graphics Society (CGS). His work on telehealth has been recognized with multiple awards, including the Health Secretary Innovation Award at the NSW Health Innovation Symposium in 2016.
\end{IEEEbiography}

\begin{IEEEbiography}[{\includegraphics[width=1in,height=1.25in,clip,keepaspectratio]{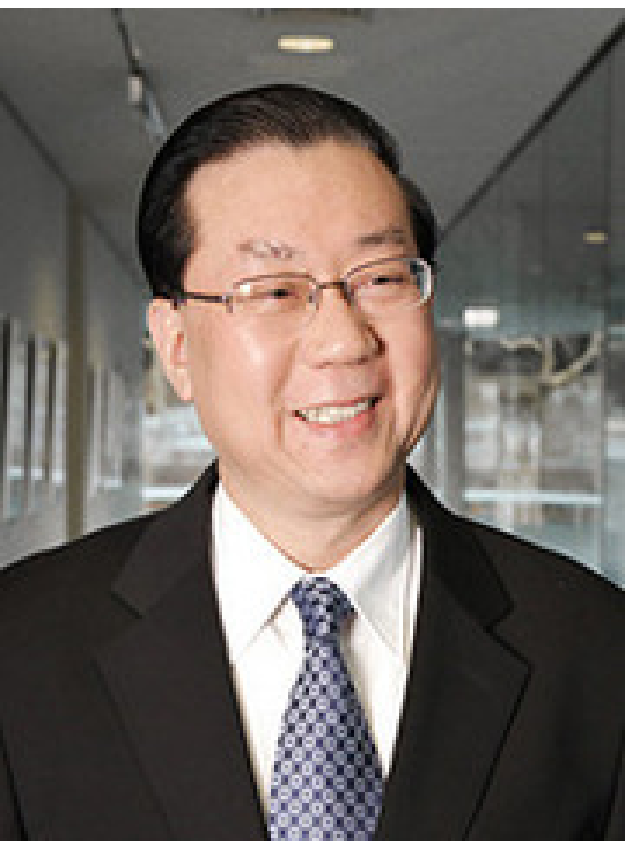}}]{David Dagan Feng}(Life Fellow, IEEE) received the M.Eng. degree in electrical engineering and computer science (EECS) from Shanghai Jiao Tong University, Shanghai, China, in 1982, and the M.Sc. degree in biocybernetics and a Ph.D. degree in computer science from the University of California, Los Angeles (UCLA), Los Angeles, CA, USA, in 1985 and 1988, respectively, where he received the Crump Prize for Excellence in Medical Engineering. He is currently the Head of the School of Computer Science, the Director of the Biomedical and Multimedia Information Technology Research Group, and the Research Director of the Institute of Biomedical Engineering and Technology, The University of Sydney, Sydney, Australia. He has published over 700 scholarly research papers, pioneered several new research directions, and made a number of landmark contributions in his field. More importantly, however, is that many of his research results have been translated into solutions to real-life problems and have made tremendous improvements to the quality of life for those concerned. He is a fellow of the Australian Academy of Technological Sciences and Engineering. He has served as the Chair for the International Federation of Automatic Control (IFAC) Technical Committee on Biological and Medical Systems, has organized/chaired over 100 major international conferences/symposia/workshops, and has been invited to give over 100 keynote presentations in 23 countries and regions.
\end{IEEEbiography}

\end{document}